\documentclass[sn-mathphys-num]{sn-jnl}

\usepackage[title]{appendix} 
\usepackage{rotating} 
\usepackage{booktabs} 
\usepackage{adjustbox} 
\usepackage{subcaption} 

\usepackage{graphicx} 
\usepackage{tikz} 

\usepackage{multirow} 

\usepackage{amsmath,amssymb,amsfonts} 
\usepackage{amsthm} 
\usepackage{mathrsfs} 
\usepackage{bm} 
\usepackage{bbm} 

\usepackage{algorithm} 
\usepackage{algorithmicx} 
\usepackage{algpseudocode} 

\usepackage{natbib} 

\usepackage{xcolor} 
\usepackage{textcomp} 
\usepackage{lipsum} 
\usepackage{listings} 
\usepackage{easyReview} 
\usepackage{manyfoot} 
\usepackage{setspace} 

\begin{document}
	
	
	\title{Goodness of fit of relational event models}
	
	\author*[1]{\fnm{Martina} \sur{Boschi}}\email{martina.boschi@usi.ch}
	
	\author[2]{\fnm{Ernst Jan Camiel} \sur{Wit}}\email{ernst.jan.camiel.wit@usi.ch}
	
	\affil*[1,2]{\orgdiv{Faculty of Informatics}, \orgname{Universit\`a della Svizzera italiana}, \orgaddress{\street{Via Giuseppe Buffi, 13}, \city{Lugano}, \postcode{6900}, \country{Switzerland}}}
	
	\abstract{A type of dynamic network involves temporally ordered interactions between actors, where past network configurations may influence future ones. The relational event model can be used to identify the underlying dynamics that drive interactions among system components. Despite the rapid development of this model over the past 15 years, an ongoing area of research revolves around evaluating the goodness of fit of this model, especially when it incorporates time-varying and random effects. Current methodologies often rely on comparing observed and simulated events using specific statistics, but this can be computationally intensive, and requires various assumptions.
		
	We propose an additive mixed-effect relational event model estimated via case-control sampling, and introduce a versatile framework for testing the goodness of fit of such models using weighted martingale residuals. Our focus is on a Kolmogorov-Smirnov type test designed to assess if covariates are accurately modeled. Our approach can be easily extended to evaluate whether other features of network dynamics have been appropriately incorporated into the model. We assess the goodness of fit of various relational event models using synthetic data to evaluate the test's power and coverage. Furthermore, we apply the method to a social study involving 57,791 emails sent by 159 employees of a Polish manufacturing company in 2010.
		
	The method is implemented in the R package \textsf{mgcv}.}
	
	
	\maketitle
	
	\section{Introduction}\label{sec:introduction}
		
	
	Many real world processes can be described as dynamic networks, where edges appear ordered in time. The vertices in these networks can be social actors, biological species, geographical regions, hospital structures, or scientific patents, depending on the context \citep{uzaheta2023, Juozaitiene2023, vu2017, filippi2023a}. And although the type of relations can vary from social interactions, invasions of geographical regions by non-native species, patient transfers between hospitals, or patent citations, edges are characterized by time-specific interactions between these vertices.  
		
	Over the past 15 years, Relational Event Models (REMs) have emerged as a useful framework for studying the dynamics that drive these events. The initial formulations of REMs within the domain of social analysis focused on modelling how simple endogenous network statistics, such as reciprocity and transitivity, influence the occurrence of these relational events \citep{butts2008}. The counting process model formulation provided a principled mathematical description of the relational event process \citep{perry2013}. Recent more flexible REM formulations accommodate time-varying and random effects, with concerted efforts to alleviate the computational cost associated with REM inference techniques \citep{boschi2023, filippi2023b}. Furthermore, \cite{lerner2022} proposed an extension to REMs to incorporate the collective participation of actors in a shared event, giving rise to the concept of relational hyper-event models.
		
	Despite the progress in relational event modelling, the contentious issue of evaluating their fit to the data persists \citep{bianchi2023}. \cite{brandenberger2019} proposed a simulation-based approach that involves simulating a segment of the event sequence and comparing the simulated events with the actual relational events. This method, while comprehensive, is time-consuming, computationally expensive, and relies on several assumptions and comparison metrics. Recently, \cite{amati2024goodness} introduced another simulation-based procedure that compares several non-modelled, relevant statistics in both real and simulated data based on the fitted model. The drawback of both approaches is that simulation-based approaches in relational event modeling are computationally very intensive, as it necessitates calculating endogenous statistics at each time point for all potential dyads at risk of interacting. Alternatively, conventional Cox regression approaches, including the assessment of Schoenfield, Deviance, and Martingale residuals, have been adapted for REMs \citep{Juozaitiene2023}. However, these techniques lack formalization, and have not been extended to the latest REM extensions mentioned above.
	
	This paper will introduce a versatile technique for assessing the goodness of fit (GOF) of arbitrary components of any relational event model. To achieve this, we adapt a method from the literature on Cox regression \citep{lin1993, marzec1997a, marzec1997b, marzec1998, borgan2015}. Broadly speaking, this approach involves comparing an observed weighted Martingale-type process over the time window of interest with its theoretical behavior under a GOF assumption. The technique seeks to identify the discrepancy between the observed statistic and its expected value under the assumed model at each time point. This sequence is then accumulated over the event sequence to produce the martingale-type process of interest  \citep{marzec1997a}. Extending the existing literature, our method is also able to test complex effects, including non-linear, time-varying, and random effects, as well as a global assessment of a model's adequacy. The technique can be generalized even further to test if any particular feature or auxiliary statistic of the system, not directly included in the model formulation, has been properly accounted for by the model. Furthermore, the method is computationally straightforward, avoiding the need for simulating relational events based on the fitted model as required by simulation-based approaches.
		
	The paper is structured as follows. In Section~\ref{sec:REMs}, we introduce a generalized additive relational event model, including non-linear and random effects, as well as a computationally efficient inference technique. Section~\ref{sec:GOF} represents the methodological core of the paper, where we describe the GOF method. For expository purposes, we introduce the GOF test in the setting of a single fixed effect, after which we show various extensions, including an omnibus test and a brief discussion on the use of statistics that are not part of the initial model formulation. In Section~\ref{sec:simulation}, we present a simulation study showing the performance of the GOF test, evaluating its power and coverage. Finally, in Section~\ref{sec:application} we apply the GOF test to a study involving a sequence of 57,791 emails sent by 159 employees from January 2nd to September 30th, 2010, within the context of a manufacturing company located in Poland \citep{michalski2011}.
		
	The datasets generated and analysed during the current study are available in the \texttt{GOF} GitHub repository  \href{https://github.com/martinaboschi/GOF.git}{here} \citep{github}. The repository also contains the \texttt{R} code used for implementing these analyses.
	
	\section{Relational event modelling}\label{sec:REMs}
	
	
	
	A \emph{relational event} denotes an interaction occurring at time $t \geq 0$, originating from a sender $s$ and directed towards a receiver $r$. Formally, it is represented as a triplet $(s, r, t)$, where $s \in \mathcal{S}$ (the set of possible senders) and $r \in \mathcal{C}$ (the set of possible receivers). Relational events can be conceptualized as the manifestations of a \emph{marked point process}, $\mathcal{M}=\{(t_k, (s_k, r_k)); k \ge 1\}$. This temporal process contains at times $t_k$ (\emph{point}) when the interaction $(s_k,r_k)$ (\emph{mark}) occurs.
	A \emph{counting process} can be associated with this marked point process, representing for each mark $(s,r)$ at any time $t\geq 0$ the number of occurred interactions $ (s, r) $ in $[0,t]$,
		\begin{equation}\label{eq:counting}
			N_{sr}(t) = \sum_{t_k \le t} \mathbbm{1}_{\{ (s_k, r_k) = (s,r)\}}, 
		\end{equation}
	In this paper, we make the following assumptions:
		\begin{enumerate}
			\item \label{item-ass1} $N_{sr}(0)=0$, meaning that no event is assumed to occur at time $0$.
			\item \label{item-ass2} $N_{sr}$ is adapted with respect to the increasing \textit{filtration} $ \mathcal{H} = \{ \mathcal{H}_t \}_{t \ge 0} $, where $\mathcal{H}_t$ is a sub-sigma field generated by the events that have occurred up to time $ t $.  
			\item  \label{item-ass3} No simultaneous events occur.
			\item  \label{item-ass4} $\forall t \in\mathbb{R}: N_{sr}(t) < \infty$ almost surely.
		\end{enumerate}
	Under these assumptions, $N_{sr}$ is a right-continuous submartingale, and, according to the Doob-Meyer theorem \citep{perry2013}, it can be decomposed as follows, 
		\begin{equation}\label{eq:doob-dec}
			M_{sr}(t) = N_{sr}(t)  - \Lambda_{sr}(t) , 
		\end{equation}
	where $ M_{sr}(t) $ is a zero-mean Martingale, and  $\Lambda_{sr}(t)=\int_0^t \lambda_{sr}(u) \mathrm{d}u$ represents the $\mathcal{H}_{t^-}$-measurable, \textit{predictable} part of the counting process $N_{sr}(t)$.
	
	\subsection{Stratified additive mixed effect REM formulation}
	
	REM involves modelling the \emph{intensity function} $ \lambda_{sr}(t) $, measuring the instantaneous hazard of the relational event $(s,r)$ at time $t$. In this paper, we consider a rather general formulation of the hazard, involving a stratified, additive, mixed effects structure. It considers a stratified formulation of $ \lambda_{sr}(t) $ that depends on $p$ potentially time-varying covariates with fixed linear (FLE), or time-varying effects (TVE), $w$ covariates with non-linear effects (NLE), and $c$ random effects (RE),
		\begin{eqnarray}
			\lambda_{sr}(t) &= Y_{sr}(t)\lambda_{0g(s,r)}(t) \exp{\left[ \bm{\beta}(t)^T \bm{x}_{sr}(t) + \bm{f}^T[\bm{v}_{sr}(t)]\mathbbm{1}_w+ \bm{b}^T \bm{z}_{sr}(t)\right]} \label{eq:model-compl} \\
			&\bm{b} \sim \mathcal{N}(\bm{0}, \Sigma(\bm{\theta}_0)) \label{eq:model-re}  
		\end{eqnarray}
	In this model formulation, several key components are involved. First, $Y_{sr}(t)$ is the \textit{at-risk indicator} of event $(s,r)$ at time $t$. Then $ \lambda_{0g}(t) $ represents a non-negative \emph{baseline intensity function}, potentially specific to different \textit{population subgroups}, indexed by $g(s,r)\in \{1,\ldots,G\}$. Relational event applications offer a rich set of possibilities for stratified formulations, which may involve the sender of the relational event \citep{perry2013, bianchi2023ties}, particular population subgroups \citep{boschi2023}, or the type of event \citep{juozaitiene2022}. In this paper, we do not explicitly focus on stratification, and for this reason we will use the notation $\lambda_{0}$ instead of $\lambda_{0g(s,r)}$, but we will outline situations in which the distinction among the two is actually relevant.
	
	The \emph{covariate processes} $ \bm{x}_{sr} \in \mathbb{R}^p $, $ \bm{v}_{sr}\in \mathbb{R}^w $, and $ \bm{z}_{sr} \in \mathbb{R}^c$ are locally bounded, left-continuous, adapted to $\mathcal{H}$, and thus predictable. Covariates that are included in REMs may be \emph{exogenous}, consisting of attributes external to the system of events, of one or both actors involved in the relational event. Or they are \emph{endogenous}, incorporating the network dynamics arising from the past.  Based on their assumed impact on the intensity function (FLE, TVE, NLE, or RE), the three sets of covariates  $ \bm{x}_{sr} $, $ \bm{v}_{sr} $, and $ \bm{z}_{sr} $ are identified in their notation.  The set of $w$ variables included in the vector $ \bm{v}_{sr} $ are assumed to have \textit{non-linear effect}, namely different values have different impacts on the hazard function. The set of $w$ functions contained in $\bm{f}$ are assumed smooth functions. We also include the assumption that $\mathbb{E}[f_d(v_{d})]$ is equal to zero to avoid identifiability issues \citep{hastie1990generalized}. The main aim of this paper is to devise a generic test for checking whether the covariates $ \bm{x}_{sr} $, $ \bm{v}_{sr} $, and $ \bm{z}_{sr} $ have been included in the model in a way that is consistent with the data. 
	
	For inference purposes, we will express non-linear components as linear combinations of some functional basis $\bm{\psi}$, 
		\begin{equation}\label{eq:NLE}
			f_d(v) = \sum_{l=1}^q \delta_{dl} \cdot \psi_l(v).
		\end{equation}
	Although the bases for the various covariates are not necessarily the same, we suppress in the notation of the basis functions their dependence on the covariate. Typical bases are the cubic spline basis, cyclic regression basis and the P-spline basis. 
	
	We assume that $p_0\le p$ covariates in $\bm{x}$ have a FLE: in this case $\beta$ is a constant. For the remaining $p-p_0$ covariates $ \bm{\beta}$ is a  time-varying function. Typically, \textit{time-varying effect} is represented as a smooth function of time, such as a thin plate spline, which can be expressed in terms of a linear basis expansion \citep{wood17}:
		\begin{equation}\label{eq:TVE-spline}
			\beta_d(t) = \sum_{l=1}^q \beta_{dl} \cdot \psi(||t - c_{l}||) \qquad \psi(t) = t^2 \log{(t)}
		\end{equation}
	where $d>p_0$ and $c_l$ $(l=1,\ldots,q)$ consist of the set of control points. Other bases can be selected as well. Applications of these two methods can be found in \cite{Juozaitiene2023} and \cite{boschi2023}, respectively.   
		
	Finally, $ \bm{b} $ consists of the \textit{random effects}, implemented as Gaussian frailties. Random effects offer a versatile means of accounting for heterogeneity in relational dynamics in the absence of specific covariates. For instance, it can account for intrinsic heterogeneity in sender activity or receiver popularity, instead of using viral factors, such as the sender's out-degree or the receiver's in-degree \citep{juozaitiene2024nodal}. Furthermore, random effects enhance the flexibility of covariate effects, allowing for effects that may be, in a certain amount, fixed across all statistical units but also specific according to particular features within the system. 
		
	As previously mentioned, the model formulation in \eqref{eq:model-compl} is complex, yet highly modular. Simpler model formulations can be implemented by appropriately selecting from the aforementioned model components. 
		
	
	\subsection{Case-control partial likelihood}\label{subsec:inference}
	
	Let $ \mathcal{D} $ be a temporally ordered sequence composed of $ n $ relational events denoted as $ e_k = (s_k, r_k, t_k) $, where $ k = 1, \ldots, n $ according to \eqref{eq:model-compl} and \eqref{eq:model-re}. For identifiability, we assume the process starts at time $t=0$ and the last observation occurs at time $\tau = t_n$.  Under assumptions \ref{item-ass1}-\ref{item-ass4} and conditional on the random effects, the \textit{partial likelihood} (PL) is the joint product of the multinomial conditional probabilities of observing events in $\mathcal{D}$ at the time points in which the events actually occurred \citep{butts2008,perry2013}. Each of these probabilities is a ratio of the hazard of the dyad experiencing the event at time $t_k$ ($k=1, \ldots, n$), and the sum of the hazard rates for all the dyads at risk of interacting at that time, namely those composing the \emph{risk set} $\mathcal{R}_{t_k}$. The computational cost of PL is expensive, as the risk set scales with the number of possible dyads, namely $n_\mathcal{S} \times n_\mathcal{C}$. 
	
	For this reason, \emph{nested case-control} (NCC) sampling \citep{borgan1995} has been adapted to REMs \citep{vu2015, lerner2020reliability}, evaluating, at each time-point, a sample of the individuals at risk, called the \emph{sampled risk set}. The sampled risk set includes the dyad involved in the event (the \emph{case}) and $m-1$ sampled \emph{controls} (dyads that could interact but did not). A new marked point process is thus driving the relational events, $ \{\left(t_k, ((s_k, r_k), SR_k)\right); k \ge 1 \},
	$ where $SR_k \subset \mathcal{R}_{t_k}$ is the sampled risk set at time $t_k$. The marked space of this new marked point process can be defined as $
			E = \{ ((s,r), SR): s \in \mathcal{S}, r \in \mathcal{C}, SR \in \mathcal{P}_{sr} \} $.
	Here, $ \mathcal{P} $ is the power set of all possible pairs of senders and receivers in $\mathcal{S}$ and $\mathcal{C}$, respectively. The set $ \mathcal{P}_{sr} \subset \mathcal{P} $ consists of of all elements in $\mathcal{P}$ of size $m$ that include $(s,r)$. This new point process is adapted to a new filtration $\{ \mathcal{F}_t \}_{t \ge 0}, \mathcal{F}_t = \mathcal{H}_t \cup \sigma(SR_k ; t_k \le t)$, accounting for previously observed sampled risk sets. Under the assumption of \emph{independent sampling}, $ \mathcal{F}_t $-intensity processes of $ N_{sr} $ are the same as $ \mathcal{H}_t $-intensity processes. In practice, sampled controls have the same failure risk as non-sampled ones \citep{borgan1995}. As before, a counting process is associated, numbering the observed mark and the associated sampled risk set, 
		\begin{equation*}\label{eq:counting-new}
			N_{sr,SR}(t) = \sum_{t_k \le t} \mathbbm{1}\{(s_k, r_k)=(s,r), SR_k=SR\}.
		\end{equation*}
	The original counting process can be retrieved as $N_{sr} = \sum_{SR \in \mathcal{P}_{sr}} N_{sr,SR}$. The intensity process $\lambda_{sr,SR}(t) = \lambda_{sr}(t) \cdot \pi_t(SR|(s,r))$ depends on the probability of sampling the risk set $SR$ given the interacting dyad $(s,r)$. 
	
	The nested case-control sampled extension of the original relational event data $\mathcal{D}$ is indicated as $\mathcal{E}= \{ \left(t_k,s_k,r_k,SR_k\right); k=1,\ldots,n \}$, with an associated NCC partial likelihood conditional on the random effects $\bm{b}$,
		\begin{equation}\label{eq:PLsampled}
			\mathcal{L}^{PS}(\bm{\beta}, \bm{\delta},\bm{b}|\mathcal{E}) 
			= \prod_{k=1}^{n} \dfrac{\lambda_{s_kr_k}\left(t_k\right)\cdot \pi_{t_k}(SR_k|(s_k,r_k))}{\sum_{(s,r) \in SR_k} \lambda_{sr}\left(t_k\right)\cdot \pi_{t_k}(SR_k|(s,r))}
		\end{equation}
	where $SR_k$ represents a potentially complete or NCC-sampled risk set at time $t_k$. When the sampled risk set coincides with the entire risk set $SR_k=\mathcal{R}_{t_k}$, then $\mathcal{F}_{t}$ coincides with $\mathcal{H}_{t}$. 
	
	This NCC partial likelihood formulation is conditioned on the random effects. A traditional way to deal with random effects is by estimating the variance component using \textit{restricted maximum likelihood estimation} \citep{pinheiro2006mixed}. This involves integrating out $\bm{\beta}$, and $\bm{\delta}$ from the nested case-control partial likelihood. By replacing the estimates of $\bm{\theta}$, $\bm{\delta}$, and $\bm{\beta}$, the random effects $\bm{b}$ can be predicted based on their conditional expectations using the obtained estimates. 
	
	Alternatively, random effects can be included directly in the likelihood as 0-dimensional smooths \citep{wood17}. As the random effects are assumed to follow a normal distribution, their probability density a corresponds to a quadratic penalization term involving an identity penalty matrix with dimensions equal to the number of levels that the considered random factor may assume. We will focus on this latter approach because it allows us to treat random effects similarly to the other smooths involved in the model formulation.
	
	\subsection{Penalized maximum NCC partial likelihood}\label{subsec:MLE}
	
	The spline formulation for the time-varying effects, the non-linear effects, and the random effects enables to write the conditional NCC partial log-likelihood in an \textit{additive form},  	
		\begin{eqnarray*}
			\ell^{PS}(\bm{\gamma}|\mathcal{E}) 
			&=& \sum_{k=1}^{n} \log\left[ 
			\dfrac{\exp\left(\bm{\gamma}^\top \bm{h}_{s_kr_k}(t_k)\right)\cdot \pi_{t_k}(SR_k|(s_k,r_k))}{\sum_{(s,r) \in SR_k} \exp\left(\bm{\gamma}^\top \bm{h}_{sr}(t_k)\right)\cdot \pi_{t_k}(SR_k|(s,r))}
			\right],		
		\end{eqnarray*}
	where the vector $\bm{\gamma}=(\bm{\beta},\bm{\delta},\bm{b})$ collects all parameters, and the covariates and spline bases are collected in a vector  $\bm{h}_{sr}(t)= (\bm{h}^{(1)}_{sr}(t), \bm{h}^{(2)}_{sr}(t), \bm{h}^{(3)}_{sr}(t))$, where 
	\begin{equation*}
		\begin{array}{rcll}
			{h}^{(1a)}_{d,sr}(t) &=& x_{d,sr}(t) & \mbox{$x_d$ has a FLE}, d \leq p_0 \\
			{h}^{(1b)}_{q(d-p_0-1)  + l,sr}(t) &=& x_{d,sr}(t) \psi(|t-c_{l}|) & \mbox{$x_d$ has a TVE, $d \leq p_0$}, l=1,\ldots,q \\
			{h}^{(2)}_{q(d-1)  + l,sr}(t) &=& \psi_l(v_{d,sr}(t)) & \mbox{$v_d$ has a NLE}, l=1,\ldots,q \\
			{h}^{(3)}_{d,sr}(t) &=& z_{d,sr}(t) & \mbox{$z_d$ has a RE}
		\end{array}
	\end{equation*}  
	Both $\bm{\gamma}$ and $\bm{h}_{sr}(t)$ are elements of $\mathbb{R}^P$, where $P$ is the sum of $p_0 + (p-p_0) \times q$, corresponding to the fixed or time-varying effects, $w \times q$, corresponding to the non-linear effects, and $c$, corresponding to the random effects. If some of the covariates have a FLE, they do not have $q$ elements in the design matrix but only one. This formulation specifically refers to non-stratified models. For stratified models, the sampled non-event can be sampled from the same stratum $g(s,r)$ as the event $(s,r,t)$. This cancels the nuisance parameter $\lambda_{0g(s,r)}$ from both numerator and denominator, resulting in the same expression $\ell^{PS}$ above.
	
	To impose smoothness on the time-varying and non-linear terms, as well as to impose a normal prior on the random effect terms, we associate a \emph{penalty term} with each of the smooth and random terms. For more details on penalty terms, the reader is referred to \cite{wood03, wood17}. The \emph{penalized maximum likelihood estimator} (PMLE) \citep{oelker2015} maximizes the NCC log-likelihood, which is given as:
		\begin{eqnarray*} 
			\ell_\lambda^{PS}(\bm{\gamma}) &=& \ell^{PS}(\bm{\gamma})  - P_\lambda(\bm{\gamma}),
		\end{eqnarray*}
	where $P_\lambda(\bm{\gamma})  = \sum_{l=1}^{L} \lambda_l \cdot p_l(\bm{\gamma}_{\bm{i}_l})$, $L$ is the number of penalized terms, $p_l$ is the $l$th penalty function, and $\lambda_l$ the $l$ \textit{tuning parameter}. Vector $\bm{i}_l \subseteq \{1,...,P\}$ contains the indices of vector $\bm{\gamma}$ identifying the $l$th penalized parameter. 
	To find the PMLE $\hat{\bm{\gamma}}$, the penalized score is set equal to zero, 
	$\nabla \ell_\lambda^{PS}(\widehat{\bm\gamma}) = \bm{0}$, resulting in an important expression for the penalization term, evaluated at the solution $\hat{\bm{\gamma}}$:
		\begin{equation}\label{eq:approx-penalty}
			\nabla \ell^{PS}(\hat{\bm{\gamma}}) = \nabla P_\lambda(\hat{\bm{\gamma}}).
		\end{equation}
	The tuning parameters are chosen via generalized cross-validation \cite{wood17}.
	
	\subsection{REM inference as additive logistic regression}
	Throughout the rest of this paper, we will often refer to the scenario involving case-control sampling with $m=2$, where we randomly sample only one non-event for each event, i.e., $\pi_t(SR|(s,r)) = \dfrac{1}{n(t)-1}$, where $n(t)= |\mathcal{R}_t|$ and $SR \subset \mathcal{R}_{t}$, such that $(s,r) \in SR$ and $|SR|=2$. In this case the NCC partial log-likelihood simplifies:
	\begin{eqnarray*} \ell^{PS_2}(\bm{\gamma}) 
		&=&	\sum_{k=1}^{n}  \left[\bm{\gamma}^T \Delta \bm{h}_{k} - \log{\left(1+\exp{\left(\bm{\gamma}^T \Delta \bm{h}_k\right)}\right)} \right]
	\end{eqnarray*}
	where $\Delta \bm{h}_k = \bm{h}_{s_kr_k}(t_k) - \bm{h}_{s_k^*r_k^*}(t_k)$ is the difference in the model matrix vector for the event and sampled non-event. This corresponds to the likelihood of a \textit{logistic regression} \citep{boschi2023, filippi2023b} with fixed response equal to $ 1 $, without intercept term, and with covariates that correspond to the difference of covariates of events and those of the sampled non-event. Inference via the penalized NCC partial log-likelihood $\ell_\lambda^{PS_2}$ allows for computationally efficient estimation of model parameters while maintaining consistency.  

	\section{Goodness of fit of relational event models}\label{sec:GOF}
	The \textit{goodness of fit} framework that we will introduce here is based on the idea that although the maximum likelihood is an \emph{overall} perfect fit to the data, at various points in the temporal relational event process there may be significant differences between observed statistics in the data and their expected value according to the fitted relational model. In an ideal scenario, where the model effectively captures all the relevant underlying dynamics, these disparities are expected to fluctuate around zero.
	Although our proposal is inspired by \cite{marzec1997a, marzec1997b, marzec1998, lin1993, borgan2015}, our methodology extends beyond fixed covariate assessments alone, enabling the evaluation of any generic process, presumed to be left-continuous and adapted to the filtration of the event-generating process, and to the augmented filtration incorporating sampling information. 
		
	This section is structured as follows. In section~\ref{subsec:prel-definitions}, we introduce the cumulative martingale residual process, which in its simplest form corresponds to the cumulative score process. In section~\ref{subsec:theor-behaviour} we derive the asymptotic distribution of this process. Subsequently, in section~\ref{subsec:KStest}, we adapt this process as a test statistic for assessing goodness of fit. Initially, we focus on the basic case where the statistic is a single covariate with a fixed linear effect before extending this to four additional statistical tests. In section~\ref{subsubsec:TVEandNLE}, we propose a test for covariates with time-varying or non-linear effects. Section~\ref{subsubsec:random-effects} deals with a GOF test for random effects. Section~\ref{subsec:global} considers a global test, concerning all components involved in the model matrix. Finally,  section~\ref{subsubsec:general-feature} considers the case in which the statistic of interest is not a covariate but instead consists of any feature of the relational system. In these four scenarios, the distribution of the test statistic is unknown, but can be simulated to produce empirical p-values.
		
	\subsection{Cumulative martingale residual process}\label{subsec:prel-definitions}
	Consider any $\mathcal{F}_{t^-}$-measurable \textit{statistic} $\phi_{sr}$ of interest. The aim is to check to what extent the statistic $\phi_{sr}$ is captured by the current model formulation. We define a \emph{martingale residual process} that captures the difference between the true $\phi_{sr}$ and its expected value under \eqref{eq:model-compl},
		\begin{equation}\label{eq:genprocess}
			\mathcal{G}[\bm{\gamma}, t] = \int_0^t \sum_{SR \in \mathcal{P}} \sum_{(s,r) \in SR} \left[\phi_{sr}(u) - \dfrac{\Phi^{(0)}_{SR}[\bm{\gamma},u]}{S^{(0)}_{SR}[\bm{\gamma},u]} \right] \textmd{d}N_{sr, SR}(u) \quad t \in [0,\tau]
		\end{equation}
	involving the following entities,
		\begin{equation*}
			\begin{aligned}
				&\Phi^{(0)}_{SR}[\bm{\gamma},u] = \sum_{(s,r) \in SR} \phi_{sr}(u) \cdot \exp{\left[ \bm{\gamma}^T \bm{h}_{sr}(u) \right]} \cdot \pi_{u}(SR|(s,r)), \\
				&S^{(0)}_{SR}[\bm{\gamma},u] = \sum_{(s,r) \in SR} \exp{\left[ \bm{\gamma}^T \bm{h}_{sr}(u) \right]} \cdot \pi_{u}(SR|(s,r)).\\
			\end{aligned}
		\end{equation*}
	Given an event sequence $\mathcal{E}$, including the NCC sampled risk sets, we define a weighted time-normalized observed martingale residual process for $u\in [0,1]$ as
		\begin{eqnarray}
			{G}[{\bm{\gamma}}, u|\mathcal{E}] 
			&=& \sum_{k \le \lfloor nu \rfloor}  w_{s_kr_k}(t_k) \cdot \left[ \phi_{s_kr_k}(t_k)- \dfrac{\Phi^{(0)}_{SR_k}[{\bm{\gamma}},t_k]}{S^{(0)}_{SR_k}[{\bm{\gamma}},t_k]}\right] \label{eq:G} \\
			&=& \sum_{k \le \lfloor nu \rfloor} w_{s_kr_k}(t_k) \cdot \phi_{s_kr_k}(t_k) \left[ 1 - \nabla {\Lambda}_{s_kr_k}(t_k) \right]   \label{eq:score-mart}
	\end{eqnarray}
	where $\nabla {\Lambda}_{s_kr_k}(t_k)=\exp(\gamma^\top \bm{h}_{s_kr_k}(t_k)) \cdot \pi_{t_k}(SR_k|(s_k,r_k))/\sum_{(s,r) \in SR_k} \exp{\left[ \bm{\gamma}^T \bm{h}_{sr}(t_k) \right]} \cdot \pi_{t_k}(SR_k|(s,r))$ is the $k$th increment in the cumulative intensity process due to observation $k$. This process $G$ transforms $\mathcal{G}$ to $n$ equally spaced time-points in $[0,1]$. The \textit{weights} $w_{s_kr_k}(t_k)\in [0,1]$ can be selected to give the process more efficient small sample properties, by down-weighing high-variance increments. 
	
	When the process $\phi_{sr}$ corresponds to one element of the design matrix, namely $ \phi_{sr} = h_{d, sr} $ and $w_{sr} = 1$, then \eqref{eq:G} evaluated at $u=1$ becomes
	\begin{eqnarray*}\label{eq:G-score}
		{G}[{\bm{\gamma}}, 1|\mathcal{E}] &=& \sum_{k =1}^n  \left[ h_{d, s_kr_k}(t_k)-
		\dfrac{\sum_{(s,r) \in SR_k} {h}_{d,sr} \exp{\left[ {\bm{\gamma}}^T \bm{h}_{sr}(t_k) \right]}}{\sum_{(s,r) \in SR_k} \exp{\left[ {\bm{\gamma}}^T \bm{h}_{sr}(t_k) \right]}}\right] \\
		&=& \nabla_d \ell^{PS}(\bm\gamma), \nonumber
	\end{eqnarray*}
	and is identical to the $d$th score component of the unpenalized NCC partial log-likelihood. By substituting the maximum penalized likelihood estimate $\widehat{\bm{\gamma}}$ and using \eqref{eq:approx-penalty}, we get
	\begin{equation}
		\bm{G}[\widehat{\bm{\gamma}},1|\mathcal{E}] = \nabla \ell^{PS}(\widehat{\bm\gamma}) = \nabla P_\lambda(\widehat{\bm\gamma}) .
		\label{eq:score}
	\end{equation}
	We can identify several specific cases. If $\phi_{sr}$ is a covariate $h_{d_1,sr}$ with a FLE, its corresponding penalty term is equal to zero. Therefore, the $d$th component of process \eqref{eq:G-score} returns to $0$ at the end of the observational time. This is not the case for the components of the model matrix that refer to covariates with NLE, TVE, or RE, where the penalty term is indeed different from zero. Furthermore, in such circumstances, each covariate $\bm{\phi}_{sr} =\bm{h}_{\bm{i}, sr}$ involves more than one component in the design matrix, as indicated by the indices $\bm{i}$. The associated process $\bm{G}[\hat{\bm{\gamma}}, \cdot|\mathcal{E}] $ is consequently multidimensional. 
		
	When dealing with NCC with $m=2$, the martingale residual process associated with the covariates $\phi_{sr}=\bm{h}_{sr}$ simplifies even further, 	
	\begin{equation*}
		\bm{G}[{\bm{\gamma}}, u] = \sum_{k \le \lfloor nu \rfloor} w_{s_kr_k} \cdot \left[1 - \text{logistic} \left({\bm{\gamma}}^T \Delta\bm{h}_k \right) \right] \cdot \Delta \bm{h}_k.
	\end{equation*}
		
	\subsection{Asymptotic distribution of martingale residual process}\label{subsec:theor-behaviour}
	
	As we want to use $G$ for testing purposes, we want to identify its distribution under the assumption that the model fits the data. We consider both the simpler case in which $\phi_{sr}$ is a FLE covariate, but also when $\bm{\phi}$ is multidimensional, corresponding to TVE, NLE, and RE covariates. Since the first case can be seen as a particular case of the second, we introduce here the results by referring to a multivariate process $\bm{G}[{\bm{\gamma}}, \cdot|\mathcal{E}]$. 
		
	Under the true parameter $\bm{\gamma}_0$ and mild regularity conditions, $ n^{-\frac{1}{2}} \times \bm{G}[{\bm{\gamma}}_0, \cdot] $ converges in distribution to a \textit{Gaussian process} with continuous paths with zero mean and covariance function $\mathbb{C}[\bm{G}(\bm{\gamma}_0, t), \bm{G}(\bm{\gamma}_0, u)] = \min{(t, u)} \cdot \bm{J}_{\bm{G}[\bm{\gamma}_0]}, \ t,u\in[0,1]$ \citep{zeileis2007generalized}, where $\bm{J}_{\bm{G}[\bm{\gamma}_0]} = \mathbb{V}(\bm{\phi}_{sr}(t) \left[ 1 - \nabla {\Lambda}_{sr}(t|\bm{\gamma}_0) \right])$ is defined as the variance of an individual martingale residual contribution. Whenever $ \bm{J}_{\bm{G}[\bm{\gamma}_0]} $ is non-singular, we define $\bm{W}[\bm{\gamma}_0, u] = \bm{J}_{\bm{G}[\bm{\gamma}_0]}^{-\frac{1}{2}} \times n^{-\frac{1}{2}} \times \bm{G}[{\bm{\gamma}}_0, u]$ for $u \in [0,1]$. Then, $\bm{W}[{\gamma}_0, \cdot]$ converges in distribution to a standard \textit{Brownian motion} $\bm{Z}(\cdot)$. 
		
	Given a consistent estimate $\widehat{\bm\gamma}$ and a consistent, and non-singular covariance matrix estimate $\widehat{\bm{J}}$ for $ \bm{J}_{\bm{G}[\bm{\gamma}_0]} $, we have
		\begin{equation}
			\widehat{\bm{W}}[\hat{\bm{\gamma}}, u] = \widehat{\bm{J}}^{-\frac{1}{2}} \times n^{-\frac{1}{2}} \times \bm{G}[\hat{\bm{\gamma}}, u] \qquad u \in [0,1]
		\end{equation}
	converges to a standard \textit{Brownian bridge} $\bm{Z}^0(\cdot)$, where $\bm{Z}^0(u) = \bm{Z}(u) - u\bm{Z}(1)$. 
	
	Under some regularity conditions the penalized maximum NCC partial likelihood estimate $\widehat{\gamma}$ is consistent, and some transformation of the observed information matrix $\mathcal{I}(\widehat{\gamma})/n$ is a consistent invertible estimate of the variance of process $G$ increments. This will allow us to derive an effective test-statistic for the GOF of an FLE  in section~\ref{subsec:KStest}. However, there are a number of improvements that are necessary to make $\widehat{\bm{W}}$ a viable GOF test-statistic of TVE, NLE, and RE for finite $n$, discussed in the subsequent sections.

	\subsection{GOF test for single covariate with fixed linear effect}\label{subsec:KStest}
	
	In this section, we derive a GOF statistic for the case $\phi_{sr} = x_{d,sr}$, namely $\phi_{sr}$ is a covariate that is part of the model formulation with a FLE. We recall that in this scenario ${G}[\hat{\bm{\gamma}}, u|\mathcal{E}]$ coincides with the corresponding element of the cumulative score vector, i.e. $ \nabla_d{\ell^{PS}(\hat{\bm{\gamma}})} $. The variance of individual contribution to the $d$th component of the score process is given as the $d$th diagonal element of the Fisher information matrix $\mathcal{I}[{\bm{\gamma}}]_{d,d}$ divided by $n$, which can be estimated via the observed information matrix, $\widehat{J} = \frac{\mathcal{I}[\hat{\bm{\gamma}}]_{d,d}}{n}$. 
	
	Therefore, $\widehat{W}[\hat{\bm{\gamma}}, \cdot] = {G}[\hat{\bm{\gamma}}, \cdot|\mathcal{E}] \times \sqrt{\mathcal{I}[\hat{\bm{\gamma}}]_{d,d}}$ under the null hypothesis of a correctly specified effect converges to a standard unidimensional Brownian bridge $Z^0(\cdot)$. Deviations from the null can be tested by evaluating, for example, the largest value of $\widehat{W}[\hat{\bm{\gamma}}, \cdot]$, 
	\begin{equation} 
		\label{eq:KS} 
		T_x = \sup_{u \in [0,1]} \lvert \widehat{W}[\hat{\bm{\gamma}}, u]\rvert 
	\end{equation}
	The distribution of the supremum of a Brownian Bridge has a \textit{Kolmogorov distribution}. Given the distribution of the univariate test statistic $T_x$ under the null hypothesis of interest, i.e., the adequacy of the chosen model, we can compute \textit{exact value} for the p-value when observing $T_x=t_x$,
	\begin{equation*}
		\mbox{p-value}(t_x) = 2\sum_{k=1}^\infty (-1)^{k-1} e^{-2k^2t_x^2}.
	\end{equation*}
		
	\subsection{GOF test for covariates with non-linear/time-varying effect}\label{subsubsec:TVEandNLE}
		
	When testing for the goodness of fit of a covariate with a non-linear or time-varying effect, we consider the statistics we introduced in section~\ref{subsec:MLE}. Let $\bm{i}_d$ correspond to the indices in the model matrix that represent the $d$th covariate of interest. In particular, for time-varying covariates we included the $q$-dimensional vector $\bm{h}_{\bm{i}_d,sr}(t) = \bm{\psi}(t)x_{d,sr}(t)$ in the model matrix, whereas for non-linear effects this corresponded to $\bm{h}_{\bm{i}_d,sr}(t) = \bm{\psi}(v_{d,sr}(t))$.   
	
	We consider the multivariate martingale residual process $\bm{G}[\hat{\bm{\gamma}},\cdot]$ for $\bm{\phi}_{sr}=\bm{h}_{\bm{i},sr}$. For finite $n$, the process $\bm{G}[\hat{\bm{\gamma}},\cdot]$ arrives at $\nabla_{\bm{i}} \ell^{PS}(\hat{\bm{\gamma}})$. To improve the Brownian bridge approximation, we define a centered process by subtracting, for each term, a fraction of the penalty term,
		\begin{equation}\label{eq:ind-score-corrected}
		\bm{G}_{k}[\hat{\bm{\gamma}}, t_k] = \left[ \bm{h}_{\bm{i}, s_kr_k}(t_k)-
		\dfrac{\sum_{(s,r) \in SR_k} \bm{h}_{\bm{i},sr} \exp{\left[ \hat{\bm{\gamma}}^T \bm{h}_{sr}(t_k) \right]}}{\sum_{(s,r) \in SR_k} \exp{\left[\hat{\bm{\gamma}}^T \bm{h}_{sr}(t_k) \right]}}\right] - \dfrac{\nabla_{\bm{i}} \ell^{PS}(\hat{\bm{\gamma}})}{n}
	\end{equation}
	Since $\bm{J}$ is defined as the variance of an individual contribution $ \bm{G}_{k}[{\bm{\gamma}}_0] $, a candidate for its estimation is the empirical variance-covariance matrix at the end of the observational period, namely $\widehat{\bm{J}} = n^{-1} \times \sum_{k=1}^n {{\bm{G}}}_{k}[\hat{\bm{\gamma}}, t_k] {{\bm{G}}}_{k}[\hat{\bm{\gamma}}, t_k]^T$ \citep{zeileis2007generalized}. These two elements now allow us to define a normalized process for $u \in [0,1]$ if the matrix $\widehat{\bm{J}}$ is invertible,
	\begin{equation}
		\label{eq:W}
		\widehat{\bm{W}}[\hat{\bm{\gamma}}, u] = \widehat{\bm{J}}^{-\frac{1}{2}} \times n^{-\frac{1}{2}} \times \sum_{k \le \lfloor nu \rfloor} \bm{G}_{k}[\hat{\bm{\gamma}}, t_k|\mathcal{E}].
	\end{equation}
	This process starts and ends at the origin for each value of $n$, and converges to a $q$-dimensional Brownian bridge $\bm{Z}^0 = \begin{bmatrix} Z_1^0(\cdot) & \hdots & Z_q^0(\cdot) \end{bmatrix}$. Following \cite{hjort2002}, we define the following test statistic,
		\begin{equation}\label{eq:KS-mult}
			T_\psi = \sup_{u\in[0,1]} \lVert \widehat{\bm{W}}[\hat{\bm{\gamma}}, u] \rVert^2 
		\end{equation}
	There is no closed form for the theoretical distribution of the supremum of the squared norm of a multivariate Brownian bridge,  $\sup_{u\in[0,1]} \lVert {{\bm{Z}}}^0(u) \rVert^2$, but it can easily be simulated and the \textit{empirical p-value} can be computed as the fraction of statistics that are larger or equal to the observed value in \eqref{eq:KS-mult}.
	
	\subsection{GOF test for random effects}\label{subsubsec:random-effects}
		
	As we interpret random effects as $0$-dimensional splines, evaluating the goodness of fit follows a similar approach to that outlined in section~\ref{subsubsec:TVEandNLE}. However, some special considerations are necessary. Typically, random effects correspond to a set of entries $\bm{i}$ of the design matrix. By identifying the statistic of interest $\bm{h}_{\bm{i}_d,sr}= \bm{z}_{d}$, where $\bm{z}_{d}$ is the $d$th random covariate vector, we can calculate the same contributions $\bm{G}_{k}[\hat{\bm\gamma},t_k]$ as defined in \eqref{eq:ind-score-corrected}. These vectors can be combined, as in \eqref{eq:W}, to obtain  a $|\bm{i}_d|$ dimensional process $\widehat{\bm W}[\hat{\bm\gamma},\cdot]$. Again, similar to \eqref{eq:KS-mult}, we can define the test-statistic $T_z$, which under the null would be the supremum of the squared norm or a standard $|\bm{i}|$ dimensional Brownian bridge.
		
	Due to the robustness of random effect estimation, the GOF test tends to have low power for detecting minor distributional deviations. Still, as we will show in section~\ref{sec:simulation}, certain forms of misspecification, especially those involving temporal misspecification of $\bm{z}_{sr}(t)$, have adequate power and can be actually detected in moderate data settings. 
		
	\subsection{Global GOF test of the relational event model}
	\label{subsec:global}
		
	The versatility of the cumulative martingale residual approach allows for its extension to a comprehensive \emph{omnibus test}. This test aims to confirm whether the model specification is appropriate \emph{overall}. As such, one would perform this test first after fitting a model. In case it gets rejected, one would proceed to the individual tests discussed before in order to diagnose exactly what part of the model has been misspecified.  
	
	We consider a collection of $L$ covariates with a combination of possible effects (FLE, TVE, NLE or RE), each associated with a set of $\bm{i}_l$ columns in the model matrix ($l=1,\ldots,L$). For each covariate $l$ we can define the $|\bm{i}_l|$-dimensional normalized process $\widehat{\bm W}[\hat{\bm\gamma},\cdot]$, the associated test statistics $T_l$ and p-values $P_l$. There is an extensive literature on how to combine p-values. Given that the p-values are correlated, we consider the Cauchy combination test \citep{liu2020cauchy}, which is a powerful test under arbitrary dependency structures that has the additional advantage of having an analytic p-value calculation. The global omnibus test-statistic is defined as
	\begin{equation}
		\label{eq:omnibus_test}  
		T_{o}= \frac{1}{L}\sum_{l=1}^L \tan \left( \pi (0.5-P_{l})\right). 
	\end{equation}
	Given the observed test-statistic $T_o=t_o$, the p-value for this global test can be calculated explicitly as,
	\[  \mbox{p-value}(t_o) = 1/2-\arctan(t_o)/\pi.\]
		
	\subsection{GOF test for auxiliary statistics}\label{subsubsec:general-feature}
	
	A model attains additional epistemological value when it transcends its immediate formulation by predicting phenomena not explicitly included within its original parametrisation. This emergent predictive power suggests a deeper correspondence between the model and the complexities of the real world phenomenon, indicating that the model captures essential truths about the system it represents. For this reason, in addition to considering testing for the GOF of model components, in this section we consider testing for arbitrary auxiliary statistics $\phi$.  For instance, we can consider using auxiliary statistics as discussed by \cite{amati2024goodness}. These are relational summary statistics of the data, such as higher-order interactions, that provides a relevant complementary perspective and may vary depending on the context. Alternatively, we can consider a possible stratification variable of the actors or events involved in the system. We would then be interested in determining whether this stratification has been adequately accounted for by the model definition that does not explicitly include this stratification. 
	
	Given a univariate auxiliary statistic $\phi_{sr}$, we define the process $G[\hat{\bm\gamma},\cdot]$ according to \eqref{eq:G} and an associated test-statistic,
	\[ T_\phi = \sup_{u \in [0,1]}|G[\hat{\bm\gamma},u]| \]
	The main challenge is to find the distribution of $T_\phi$ under the null hypothesis. In fact, when \(\phi\) is not a model component, the distributional properties outlined in section~\ref{subsec:theor-behaviour}, related to the score process, are lost. If the model is correctly specified, \cite{borgan2015} suggests approximating the distribution of \(G[\hat{\bm{\gamma}}, u|\mathcal{E}]\) by simulating multiple replicates of the process \(G^*[\hat{\bm{\gamma}}, u|\mathcal{E}]\), which only involves sampling i.i.d. standard normal distributed $N_k, k =1,\ldots,n$,
		\begin{equation}\label{eq:sim-process}
			\begin{aligned}
				&G^*[\hat{\bm{\gamma}}, u|\mathcal{E}] =  \sum_{k \le \lfloor nu \rfloor}  \left[ \phi_{s_kr_k}(t_k) -  \dfrac{\Phi^{(0)}_{SR_k}[\hat{\bm{\gamma}},t_k]}{S^{(0)}_{SR_k}[\hat{\bm{\gamma}},t_k]}  \right]\cdot N_{k}  \\
				&-  \sum_{k \le \lfloor nu \rfloor}  \dfrac{\Phi^{(0)*}_{SR_k}[\hat{\bm{\gamma}},t_k]^T}{S^{(0)}_{SR_k}[\hat{\bm{\gamma}},t_k]} \times \mathcal{I}[\hat{\bm{\gamma}}]^{-1} \cdot  \sum_{k=1}^n  \left[ \bm{h}_{s_kr_k}(t_k) - \mathbb{E}^{U}_{SR_k}\left[\hat{\bm{\gamma}},t_k\right] \right] \cdot N_{k} 
			\end{aligned}
		\end{equation}
	where,
	\small
	\begin{equation}
		 \begin{aligned}
			 	& \Phi^{(0)*}_{SR_k}[\hat{\bm{\gamma}},t_k] = \sum_{(s,r) \in SR_k} \left( \bm{h}_{sr}(t_k) - \mathbb{E}^{U}_{SR_k} \left[\hat{\bm{\gamma}},t_k\right] \right) \cdot \phi_{sr}(t_k) \cdot \exp{\left[ \hat{\bm{\gamma}}^T \bm{h}_{sr}(t_k) \right]} \cdot \pi_{t_k}(SR_k|(s,r)) \\
			 	& S^{(1)}_{SR_k}[\hat{\bm{\gamma}},t_k] = \sum_{(s,r) \in SR_k} \bm{h}_{sr}(t_k) \cdot \exp{\left[\hat{\bm{\gamma}}^T \bm{h}_{sr}(t_k) \right]} \cdot \pi_{t_k}(SR_k|(s,r)) \\
			 	& \mathbb{E}^{U}_{SR_k}\left[\hat{\bm{\gamma}},t_k\right] = \frac{S^{(1)}_{SR_k}[\hat{\bm{\gamma}},t_k]}{S^{(0)}_{SR_k}[\hat{\bm{\gamma}},t_k]}.
			 \end{aligned}
	\end{equation}
	\normalsize
	$\mathbb{E}^{U}_{SR}\left[\hat{\bm{\gamma}},u\right]$ corresponds to the expectation of the process $\bm{h}_{sr}$ under the estimated model. The expression $\sum_{k =1}^n  \left[ \bm{h}_{s_kr_k}(t_k) - \mathbb{E}^{U}_{SR_k}\left[\hat{\bm{\gamma}},t_k\right] \right]$ is another way of writing the score \eqref{eq:score}. This approach has been shown to converge to the distribution of $G[\hat{\bm\gamma}]$. 
		
	Given $b=1,\ldots,B$ samples $G_b^*[\hat{\bm\gamma},\cdot]$ of the simulated process along with their associated test-statistic $t^*_{\phi,b}$, we can obtain an empirical p-value for $T_\phi=t_\phi$ by adapting the statistical testing procedure from \cite{borgan2015} as
	\[   \widehat{\mbox{p-value}}(t_\phi) = \frac{1}{B}\sum_{b=1}^B \mathbbm{1}\{t_{\phi,b}^* \ge t_\phi\}. \] 
		
	Alternatively, under the assumptions stated in \cite{marzec1997a}, it is possible to prove that $ n^{-\frac{1}{2}} \times G[\hat{\bm{\gamma}}, \cdot] $, even in its general formulation, weakly converges in distribution to a zero-mean Gaussian process, which leads to the possibility of a formal $\chi^2$ statistical test by using $\widehat{W}[\hat{\bm{\gamma}},1]$. 

	\section{Simulation studies}\label{sec:simulation}
		
	This section provides an empirical evaluation of the goodness-of-fit tests presented in section~\ref{sec:GOF}. We will examine both \textit{coverage} and \textit{power}, two key factors essential for determining the effectiveness of a statistical test. To achieve this, we will investigate several scenarios, explaining the test's adaptability and issues in each.
		
	Relational event occurrences are simulated according to a marked counting process described in \eqref{eq:model-compl}. Technical details can be found in the Supplementary Materials. We consider a \textit{data generation process} (DGP), which may involve linear, time-varying, non-linear, and random effects of various factors. Each simulation setting is repeated $n_\textmd{sim} = 500$ times. In order to study coverage (section~\ref{subsec:coverage}) and power (section~\ref{subsec:power}) of the testing procedure, we examine situations where the fitted model is correctly specified and where it is not, respectively. In section~\ref{subsec:sim-omnibus}, we study the omnibus test to evaluate the relational event model in its entirety.
		
	\subsection{Coverage of goodness-of-fit test}\label{subsec:coverage}
		
	When the model is correctly specified, we expect that p-values of the GOF test are uniformly distributed between $0$ and $1$. In particular, the proportion of rejections obtained by a correct statistical testing procedure is approximately equal to the \textit{significance level}, which, for simplicity, is set equal to $\alpha = 0.05$ in the simulations. 
		
	\subsubsection{Coverage of NLE GOF test with increasing sample sizes}
		
	In the first simulation study, we investigate how the coverage of the GOF test for non-linear effects evolves with increasing sample size, \( n = 1000, 5000, 10000, \) and \( 50000 \). Relational event models capture dynamics like reciprocity, common in real-world interactions such as email exchanges or trading favours. In our simulation study, we focus on reciprocity as a driving factor. The DGP incorporates reciprocity in an intrinsic non-linear way, namely as a non-linear function of the time since the last reciprocal event, which allows for a higher hazard closer to this event, which reduces over time. We estimate the non-linear effect using a $10$-dimensional thin plate regression spline, which smoothly varies the reciprocity effect over time, reflecting the flexibility that characterizes the DGP. 
	
	\begin{figure}[tb]
		\begin{tabular}{cc}
		\includegraphics[width=0.49\linewidth]{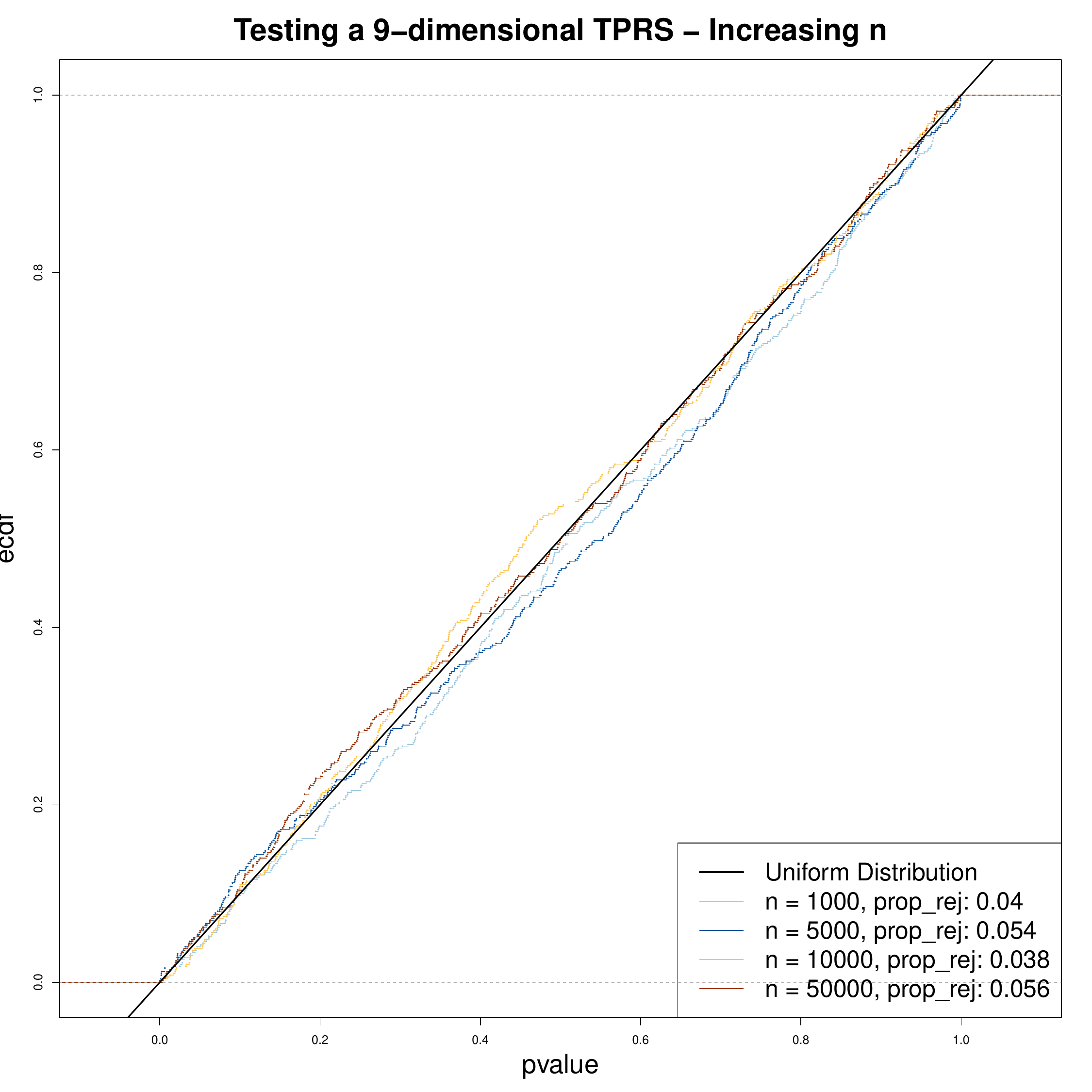} & 
		\includegraphics[width=0.49\linewidth]{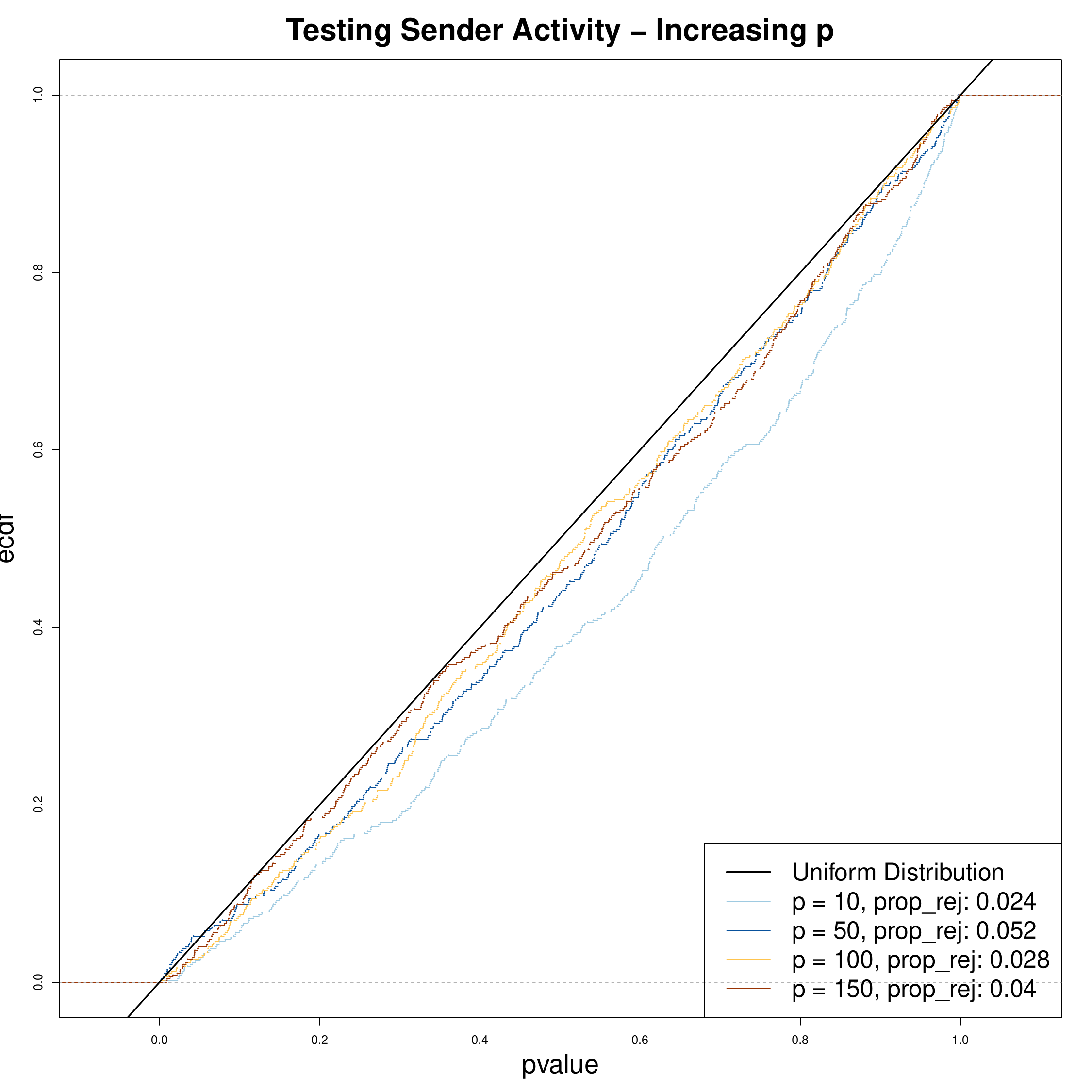}\\
		(a) & (b)
	\end{tabular}
		\caption{Coverage of the statistical procedure. \emph{Left}. Empirical p-value distribution under the null hypothesis of correctly incorporating a non-linear reciprocity effect for different sample sizes ($ n = 1000, 5000, 10000, 50000 $) with proposed GOF test in  \eqref{eq:KS-mult}. Reciprocity has been fitted by employing a $9$-dimensional thin plate regression spline. The test achieves almost perfect coverage over a wide range of sample sizes.  \emph{Right}. Empirical p-value distribution under the null hypothesis of correctly incorporating a random sender effect for different numbers of actors ($ n_\mathcal{S} = n_\mathcal{C}  = 10, 50, 100, 150 $) with the proposed GOF test in section~\ref{subsubsec:random-effects}. The test tends to be somewhat conservative for a low number of actors, whereas they appear more uniform when the number of actors is larger. \label{fig:coverage}
		}
	\end{figure}
	
	For each simulated dataset, we obtain a model fit and evaluate the multivariate observed martingale residual process $\bm{G}[\hat{\bm\gamma},\cdot]$, in \eqref{eq:ind-score-corrected} and calculate test statistics following \eqref{eq:KS-mult}. Since the model includes only reciprocity, the test involves the entire score vector. The empirical \( p \)-values are found by simulating $9$-dimensional Brownian bridges, reproducing the theoretical behaviour of the observed score process. Figure~\ref{fig:coverage}a shows the distribution of \( p \)-values across $n_\textmd{sim} =500 $ simulations for different sample sizes, relative to the expected black line representing the $U(0,1)$ distribution. We can conclude that the GOF test coverage performance is robust across different sample sizes. This is confirmed proportion of rejections that is approaching the expected value of $0.05$.
	
	\subsubsection{Coverage of RE GOF test with increasing number of actors}
	
	In this simulation study, we investigate how the coverage of the GOF test for random effects proposed in section~ref{subsubsec:random-effects} evolves with an increasing number of actors, $p = 10, 50, 100$, and $150$. In the simulation, events are exclusively driven by the sender's intrinsic properties. In other terms, the event generation is driven by an intensity function that only depends on a sender-specific parameter, that has been sampled from a Gaussian distribution. For each simulated data, we apply a model incorporating sender-related random effects, fitted as $0$-dimensional splines. This model is correctly specified, so we expect the empirical \( p \)-values to be distributed uniformly. 
		
	As Figure~\ref{fig:coverage}b shows, the goodness of fit test for random effects seems to be slightly too conservative when there is only a small number of actors in the system. As soon as $p=n_\mathcal{S} = n_\mathcal{C} $ increases, the GOF test achieves nominal coverage, as indicated by the fact that the distribution converges to a $\mathcal{U}(0,1)$ distribution. 	

	\subsection{Power of goodness-of-fit test}\label{subsec:power}
		
	The power of a statistical test is its ability to correctly reject a null hypothesis when it is false. The power of the test can be evaluated by applying it to a misspecified model.
		
	\subsubsection{Detecting FLE misspecification with increasing sample size}
	
	In this section, we evaluate how the GOF test can detect a misspecified linear effect as an increasing function of the sample size, \( n = 1000, 5000, 10000, \) and \( 50000 \). As in the previous section, we simulate a process with a non-linear function of reciprocity, defined as the inter-arrival time since the last reciprocal event occurred. In this section, we fit a misspecified model by assuming a linear effect for reciprocity. 
	
	\begin{figure}[tb]
		\begin{tabular}{cc}
			\includegraphics[width=0.49\linewidth]{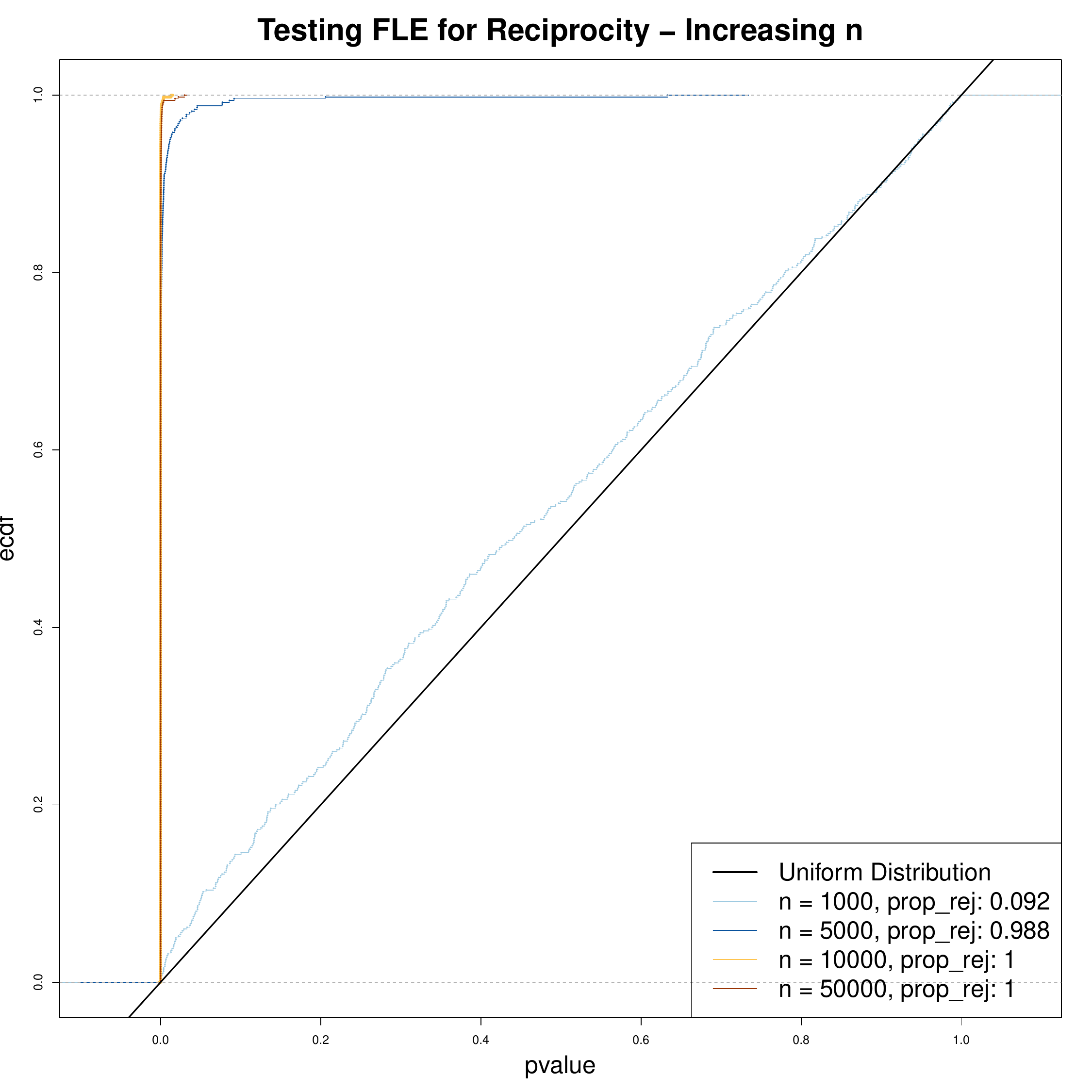} & 
			\includegraphics[width=0.49\linewidth]{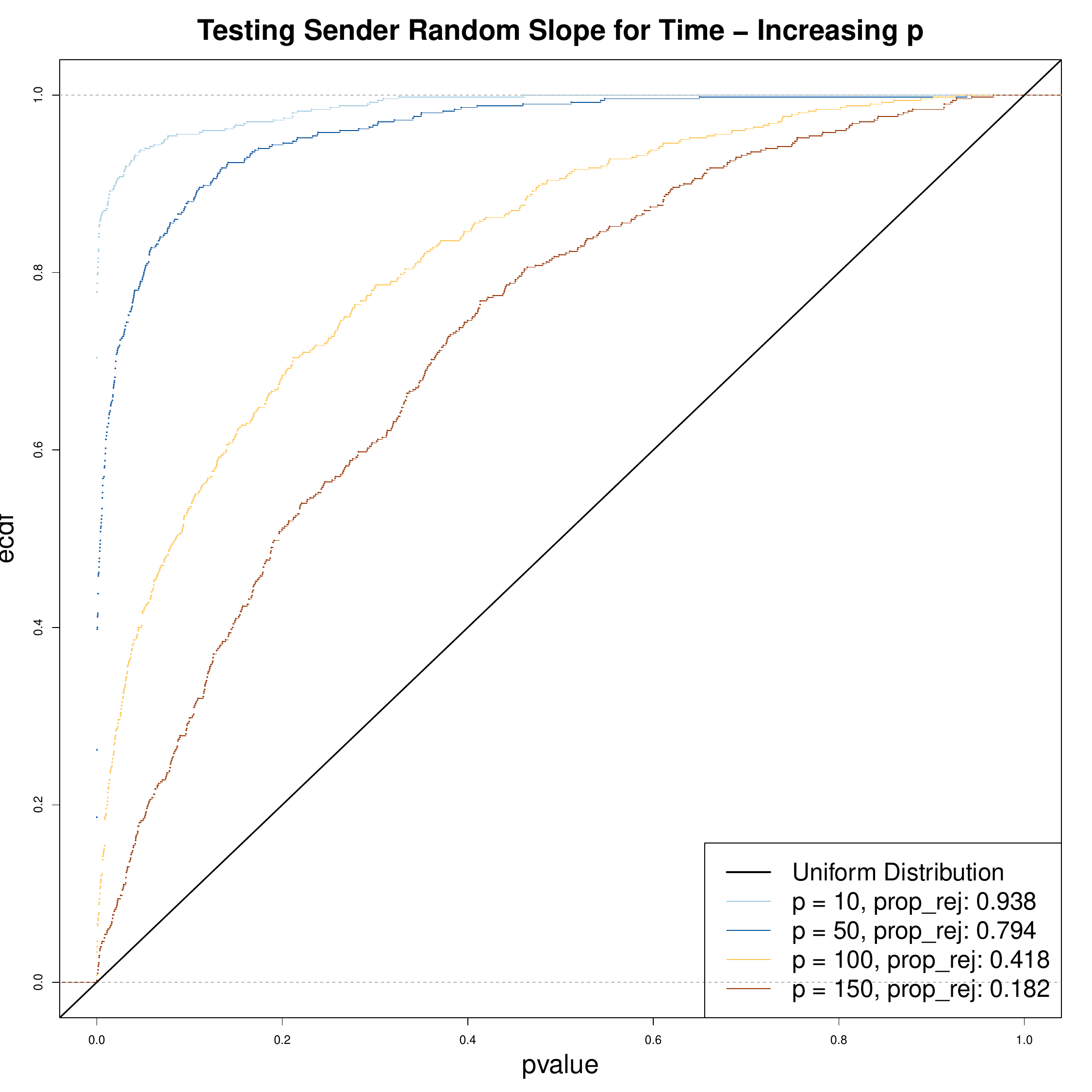}\\
			(a) & (b)
		\end{tabular}	
		\caption{Power of the statistical procedure.. \emph{Left}. Incorporating reciprocity linearly. For each size of the experiment ($ n = 1000, 5000, 10000, 50000 $), the picture reports the empirical distribution of empirical $p$-values. Since we test a covariate with a fixed effect, we can rely on  \eqref{eq:KS}. The power increases as the size increases. \emph{Right}. Events are driven by sender intrinsic properties. In this case, for each number of actors ($ n_\mathcal{S} = n_\mathcal{C}  = 10, 50, 100, 150 $), the model inadequately incorporates a different impact of time of the event according to the sender involved. Test follows  \eqref{eq:KS-mult}. Although the misspecification is clearly detected, the power tends to reduce as the number of entities increases. \label{fig:power}}
	\end{figure}
	
	Figure~\ref{fig:power}a shows a clear increase in power as the number of events increases. This allows us to conclude that the test becomes more reliable and robust with larger sample sizes. There is another issue that exacerbates this point: as the event sequence increases, the time-sequence lengthens, which will affect for a number of events the time since the last reciprocal event. This means that the linear effect for reciprocity simply becomes a worse approximation with increasing $n$. This further necessitates using a smooth function capable of accommodating the actual non-linear effect. 
	 
	\subsubsection{Detecting RE misspecification with increasing actor numbers}
	 	
	In this section, we test the ability of the goodness-of-fit test for random effects to detect misspecification for increasing network sizes, $p = 10, 50, 100$, and $150$. Again, as in the previous section, the simulation consists of a random intercept model for each of the senders of the events. Instead, the fitted model considers a random slope model for each of the senders as a function of time, incorrectly assuming that the impact of the interactions is affected by the time of the interaction.  
	
	Figure~\ref{fig:power}b shows that the misspecification in this case is clearly detected. Perhaps somewhat surprisingly, as the number of actors increases there is a reduction in the power of the test. What is happening seems to be that as the number of relational events $n$ is kept fixed in the simulations, the increase in the number of actors reduces the overall accuracy of the model fit and therefore the power of the test. Therefore, it may be that the power reduction is due to the increase in the effective number of parameters in the model, rather than to the misspecification itself. 
	
	The conducted simulation studies show that our technique has high power and flexibility, identifying several forms of misspecification. The result is confirmed in the following section, where all these components are considered in an omnibus test. 
	
	\subsection{Global testing of a relational event model}\label{subsec:sim-omnibus}
		
	We conclude with an example of the application of the global goodness-of-fit test. The omnibus test introduced in section~\ref{subsec:global} aims at confirming whether or not the model can be deemed to have an appropriate fit to the data \emph{overall}. In this simulation study, we rely on $4$ DGPs. The DGPs progressively increase in complexity, by adding one of the following additional covariates:
		\begin{enumerate}
			\item Time-based covariate for reciprocity; in this case, the reciprocity measured as defined above is assumed to have a linear impact on the intensity function; 
			\item Exogenous covariate, sampled from an exponential distribution, with a linear effect;
			\item Exogenous covariate, sampled from a Gaussian distribution, with a time-varying effect;
			\item Random intercept sender effect.
		\end{enumerate}
	So, the first DGP incorporates only reciprocity, while the last one includes all the four covariates mentioned above. 
		
	To evaluate coverage, we fit the correctly specified model for each of these four datasets and, as before, we look to the empirical distribution of the $p$-values. In this scenario, we rely on the statistical test-statistic $T_o$. We note that the first two covariates are uni-dimensional, the third one is a $9$-dimensional thin plate regression spline, and the fourth covariate is $50$-dimensional random factor. 
		
	\begin{figure}[tb]
		\begin{tabular}{cc}
			\includegraphics[width=0.49\linewidth]{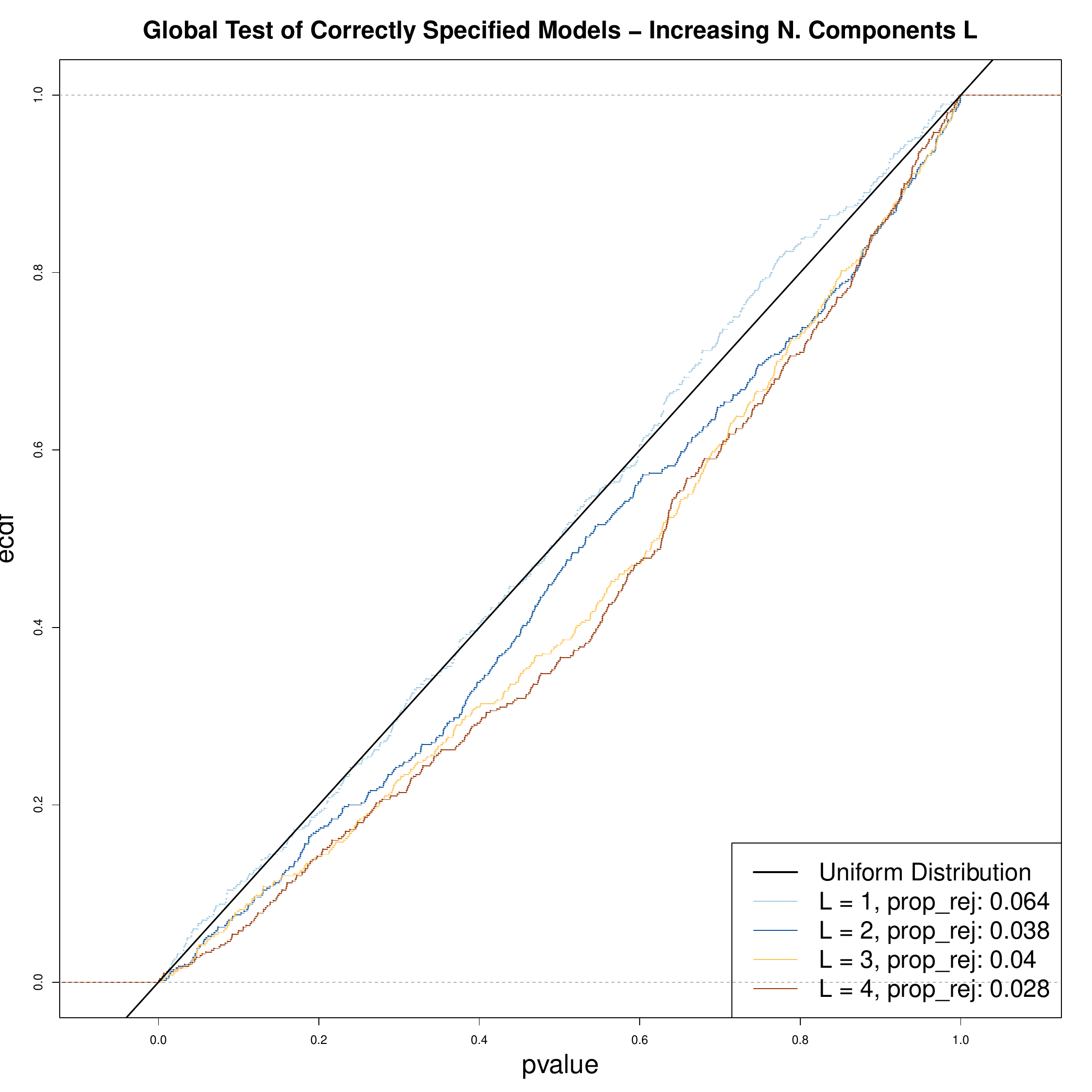} & 
			\includegraphics[width=0.49\linewidth]{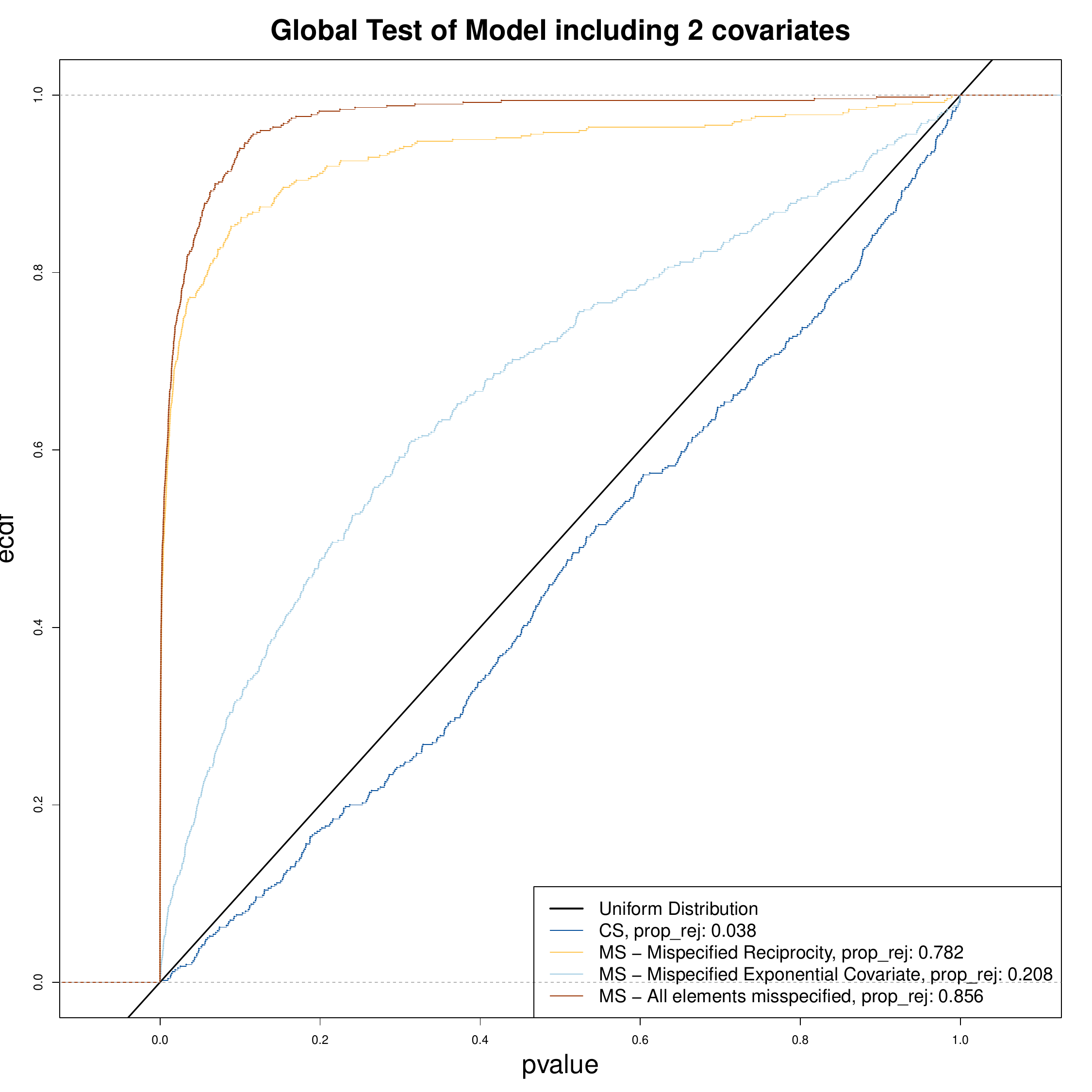}\\
			(a) & (b)
		\end{tabular}	
		\caption{Omnibus statistical procedure.. \emph{Left.} For each model including a different number of groups of elements ($ L = 1, 2, 3, 4 $), we show the empirical distribution of the simulated p-values related to the test of the overall adequacy of the relational event model according to the test in \eqref{eq:omnibus_test}. \emph{Right.} In the scenario in which $2$ elements are incorporated in the model formulation, four model are tested: the correctly specified (CS) and those with one or both elements misspecified (MS).}
			\label{fig:global}
	\end{figure}
	
	Figure~\ref{fig:global}a confirms the reliability of the omnibus test, even if we outline that, especially when including the random effect, the test tends to be slightly conservative. We think that this may be because the test makes no assumptions about the dependence between the statistics. Since it is quite common to deal with correlated statistics in REMs, this lack of assumption is relevant even if it comes at the price of a slightly conservativeness of the test.
	Figure~\ref{fig:global}b shows four possibilities for a model with $L=2$ covariates: both covariates are correctly specified, one is incorrectly specified or both are incorrectly specified. The completely misspecified model is most easily identified, whereas the models with one misspecified element have a lower power. The Supplementary materials report all other graphical results.
		
	The purpose of the global test is that once a model has been identified as misspecified, one can go further and investigate the source of the problem. For example, in the scenario in which $2$ elements are incorporated in the model formulation, it might be that only one of them or both of them are not modeled properly. In the Supplementary Materials, the reader can find for each of the fitted models both the global and the group-specific tests of adequacy. 
		
	\section{Email communication in a manufacturing company}\label{sec:application}
		
	This section examines the email exchanges among employees of a manufacturing company located in Poland \citep{michalski2011}. This dataset has previously been analyzed to investigate the temporal component of social dynamics \citep{juozaitiene2022}. The original dataset, detailed in \cite{michalski2014}, comprises over 82,000 emails sent between January 2nd, 2010, and September 30th, 2010. After a preliminary analysis aimed at removing duplicates and self-sent emails, the final dataset consists of \(n_\mathcal{E} = 57791\) relational events involving \(n_\mathcal{S} = n_\mathcal{C} = 159\) employees.
		
	\subsection{Incorporating temporal structure in relational dynamics}\label{subsec:reldyn}
		
	\citep{juozaitiene2022} focused their analysis on the temporal aspects of reciprocity, repetition, and cyclic closure. Here, we additionally include transitive closure. \textit{Reciprocity} refers to the mutual exchange of goods, services, or behaviours between actors, relying on a \textit{give and take} principle that encourages collaboration among the parties involved. In the context of email communication, this typically manifests as a tendency to respond courteously to emails. \textit{Repetition} reflects the tendency to maintain established patterns in communication. In email exchanges, this might involve following up with previously contacted counterparts. \textit{Cyclic closure} and \emph{transitive closure} are examples of triadic closure, which involves an event between two actors that have previously been in touch with a third actor. In cyclic closure the event $(s,r)$ follows the events $(r,k)$ and $(k,s)$, whereas transitive closure it follows the events $(s,k)$ and $(k,r)$. Triadic closure is known for its role in path abbreviation to enable contact between previously unconnected individuals. 
		
	Once the endogenous covariates of interest are defined, the next concern is their quantification. Table~\ref{tab:relational_dynamics} outlines two potential approaches for quantifying each endogenous covariate, highlighting different choices for assessing the presence and magnitude of these phenomena. The initial approach involves a simple identity function, determining whether the event of interest occurred before the current time. The second method incorporates a temporal dimension by analyzing the time interval between the occurrences of current time and the time of the relevant event in the past. In order to deal with the possibility that the event may not have occurred at all, we generalize the expression for the time-interval as follows:
		\begin{equation}
			\label{eq:timescale}
			\Delta \tau_{sr}^* = \exp{[-b \times (t - t^*(s,r))]}
		\end{equation}
	where \(t^*(s,r)\) represents the time of the event of interest. For example, for reciprocity this could be the last event $(r,s)$ prior to $t$, i.e., $t^*(s,r) = \max_{s_k=r,r_k=s,t_k<t} t_k$. Practically, it can be computed even when the event of interest has not occurred, yielding a value of \(0\). The tuning parameter \(b\) allows for adjustment of the time difference values to improve uniformity of the model matrix elements in the interval $[0,1]$. Although the time-scale may not be immediately interpretable, the effect can be easily transformed to the original time-scale. 
	
	Furthermore, several definitions and conceptualizations of these endogenous covariates can be proposed and must then be chosen and evaluated: first, considering which events of interest to include. For instance, when multiple reciprocal events exist in the past, we might choose the first one, the last one, or the average time of the occurred ones; in this work, we refer to the most recent one. Additionally, we may include a lifespan in the decay function of the time window. Finally, we need to decide about the closure of the mentioned dynamics: for example, in our conceptualization, responding to a reciprocal event makes that event no longer reciprocal. This is a choice and can be potentially be evaluated by means of a goodness-of-fit test. Some potential and assumed choices are described in Table~\ref{tab:relational_dynamics}.
			
		\begin{sidewaystable}[h!]
			\centering
			\footnotesize
			\begin{adjustbox}{max width=0.9\textwidth, max height= 0.9\textheight}
				\begin{tabular}{|p{0.15\textwidth}|p{0.15\textwidth}|p{0.25\textwidth}|p{0.18\textwidth}|p{0.18\textwidth}|}
					
					\hline
					\vspace{2mm} \textbf{Relational Dynamics} \vspace{4mm} & 
					\vspace{2mm} \textbf{Interpretation} \vspace{4mm} & 
					\vspace{2mm} \textbf{Representation} \vspace{4mm} & 
					\vspace{2mm} \textbf{Identity Covariate} \vspace{4mm} & 
					\vspace{2mm} \textbf{Time-Related Covariate} \vspace{4mm} \\
					
					\hline
					\vspace{2mm} \textbf{Reciprocity} \vspace{4mm} & 
					\vspace{2mm} If the contribution of reciprocity is positively estimated, then there is a tendency to send emails to counterparts who have previously initiated communication.\vspace{4mm} &
					\vspace{2mm} 
					\begin{tikzpicture}
						\node at (0,0) [left] {$s$};
						\node at (4,0) [right] {$r$};
						\node[draw=black, thick, circle, fill=lightgray, inner sep=2pt] (s) at (-0.2,0) {$s$};
						\node[draw=black, thick, circle, fill=lightgray, inner sep=2pt] (r) at (4.2,0) {$r$};
						\draw[-latex, thick, bend right=30] (s) to node[midway, above] {\(t\)} (r);
						\draw[dotted, thick, bend right=30, -latex] (r) to node[midway, above] {\(t_{rs}^l\)} (s);
					\end{tikzpicture} 
					\begin{equation*}
						t^l_{sr} < t^l_{rs} < t
					\end{equation*}
					\vspace{4mm} & 
					\vspace{2mm} 
					\begin{displaymath}
						\begin{aligned}
							&\textmd{rec}^{\textmd{id}}_{sr}(t) = \\
							&1_{ \left[t > t^l_{rs}\right]} \\
							&t^l_{sr} < t^l_{rs} < t
						\end{aligned}
					\end{displaymath}
					\vspace{4mm}& 
					\vspace{2mm}
					\begin{displaymath}
						\begin{aligned}
							& \textmd{rec}^{\textmd{time}}_{sr}(t) = \\
							& \exp{ \left[-\left(t-t^l_{rs}\right)\right]} \\
							& t^l_{sr} < t^l_{rs} < t
						\end{aligned}
					\end{displaymath}
					\vspace{4mm} \\
					
					\hline
					\vspace{2mm} \textbf{Repetition} \vspace{4mm} & 
					\vspace{2mm}  If the contribution of repetition is positively estimated, then there is a tendency to send emails to counterparts already contacted before.
					\vspace{4mm} &
					\vspace{2mm} 
					\begin{tikzpicture}
						\node at (0,0) [left] {$s$};
						\node at (4,0) [right] {$r$};
						\node[draw=black, thick, circle, fill=lightgray, inner sep=2pt] (s) at (-0.2,0) {$s$};
						\node[draw=black, thick, circle, fill=lightgray, inner sep=2pt] (r) at (4.2,0) {$r$};
						\draw[-latex, thick, bend right=30] (s) to node[midway, above] {\(t\)} (r);
						\draw[dotted, thick, bend right=-30, -latex] (s) to node[midway, above] {\(t_{sr}^l\)} (r);
					\end{tikzpicture}
					\begin{displaymath}
						t^l_{sr} < t
					\end{displaymath}
					\vspace{4mm} & 
					\vspace{2mm} 
					\begin{displaymath}
						\begin{aligned}
							& \textmd{rep}^{\textmd{id}}_{sr}(t) = \\
							& 1_{ \left[t > t^l_{sr}\right]} \\
							& t^l_{sr} < t
						\end{aligned}
					\end{displaymath}
					\vspace{4mm}& 
					\vspace{2mm}
					\begin{displaymath}
						\begin{aligned}
							& \textmd{rep}^{\textmd{time}}_{sr}(t) = \\
							& \exp{ \left[-\left(t-t^l_{sr}\right)\right]} \\
							& t^l_{sr} < t
						\end{aligned}
					\end{displaymath}
					\vspace{4mm}\\
					
					\hline
					\vspace{2mm} \textbf{Cyclic Closure} \vspace{4mm} & 
					\vspace{2mm}  If the contribution of cyclic closure is positively estimated, it indicates a tendency for a sender to correspond with counterparts who have previously initiated communication with another employee, and subsequently, that employee has communicated with the sender. \vspace{4mm} &
					\vspace{2mm} 
					\begin{tikzpicture}
						\node[draw=black, thick, circle, fill=lightgray, inner sep=2pt] (s) at (-0.2,0) {$s$};
						\node[draw=black, thick, circle, fill=lightgray, inner sep=2pt] (r) at (4.2,0) {$r$};
						\node[draw=black, thick, circle, fill=lightgray, inner sep=2pt] (k) at (2,2.25) {$k$};
						
						\draw[thick, bend right=25, -latex] (0,0) to node[midway, above] {\(t\)} (4,0);
						\draw[dotted, thick, bend right=25, -latex] (4,0) to node[midway, above] {\(t_{rk}^l\)} (2,2);
						\draw[dotted, thick, bend right=25, -latex] (2,2) to node[midway, above] {\(t_{ks}^l\)} (0,0);
					\end{tikzpicture}
					\begin{displaymath}
						t^l_{sr} < t^l_{rk} < t^l_{ks} < t
					\end{displaymath}
					\vspace{4mm} & 
					\vspace{2mm} 	
					\begin{displaymath}
						\begin{aligned}
							& \textmd{cyc}^{\textmd{id}}_{sr}(t) = \\
							& 1_{ \left[t > t^l_{ks}\right]} \\
							& t^l_{sr} < t^l_{rk} < t^l_{ks} < t \\
							& k \text{: common counterpart}
						\end{aligned}
					\end{displaymath}
					\vspace{4mm}& 
					\vspace{2mm}
					\begin{displaymath}
						\begin{aligned}
							& \textmd{cyc}^{\textmd{time}}_{sr}(t) = \\
							& \exp{ \left[-\left(t-t^l_{ks}\right)\right]} \\
							& t^l_{sr} < t^l_{rk} < t^l_{ks} < t \\
							& k \text{: common counterpart}
						\end{aligned}
					\end{displaymath}
					\vspace{4mm}\\
					
					\hline
					\vspace{2mm} \textbf{Transitive Closure} \vspace{4mm} & 
					\vspace{2mm}  If the contribution of transitive closure is positively estimated, it indicates a tendency for a sender to directly correspond with counterparts, previously indirectly contacted by means of an intermediator. 
					\vspace{4mm} &
					\vspace{2mm} 
					\begin{tikzpicture}
						\node[draw=black, thick, circle, fill=lightgray, inner sep=2pt] (s) at (-0.2,0) {$s$};
						\node[draw=black, thick, circle, fill=lightgray, inner sep=2pt] (r) at (4.2,0) {$r$};
						\node[draw=black, thick, circle, fill=lightgray, inner sep=2pt] (k) at (2,2.25) {$k$};
						
						\draw[-latex, thick, bend right=25] (0,0) to node[midway, above] {\(t\)} (4,0);
						\draw[dotted, thick, bend right=-25, -latex] (0,0) to node[midway, above] {\(t_{sk}^l\)} (2,2);
						\draw[dotted, thick, bend right=-25, -latex] (2,2) to node[midway, above] {\(t_{kr}^l\)} (4,0);
					\end{tikzpicture}
					\begin{displaymath}
						t^l_{sr} < t^l_{sk} < t^l_{kr} < t
					\end{displaymath}
					\vspace{4mm} & 
					\vspace{2mm} 	
					\begin{displaymath}
						\begin{aligned}
							& \textmd{trs}^{\textmd{id}}_{sr}(t) = \\
							& 1_{ \left[t > t^l_{kr}\right]} \\
							& t^l_{sr} < t^l_{sk} < t^l_{kr} < t \\
							& k \text{: common counterpart}
						\end{aligned}
					\end{displaymath}
					\vspace{4mm}& 
					\vspace{2mm}
					\begin{displaymath}
						\begin{aligned}
							& \textmd{trs}^{\textmd{time}}_{sr}(t) = \\
							& \exp{ \left[-\left(t-t^l_{kr}\right)\right]} \\
							& t^l_{sr} < t^l_{sk} < t^l_{kr} < t \\
							& k \text{: common counterpart}
						\end{aligned}
					\end{displaymath}
					\vspace{4mm}\\
					\hline
				\end{tabular}
			\end{adjustbox}
			\vspace{4mm}
			\caption{Table of Relational Dynamics}
			\label{tab:relational_dynamics}
		\end{sidewaystable}
		
		\normalsize
	
	\subsection{GOF testing of temporal dynamics of email communication}\label{subsec:REMapplication}
	
	The goal of this section is to determine whether and how our goodness-of-fit test allows us to understand if some of these choices concerning the effect structure of the covariates may be more or less adequate for fitting the REM to the data at hand. We adopt the notation as outlined in Table $\ref{tab:relational_dynamics}$. Our first objective is to assess the optimal relational event model that incorporates elements of reciprocity, repetition, cyclic, and transitive closure. 
		
	Suppose, for a moment, that the precise timestamp of the emails is unknown, and only their sequential order is available. Such a scenario is often encountered in practical situations \citep{butts2008}. In this case, it remains feasible to assess the endogenous dynamics for all four covariates using an identity function. The fitting procedure leads us to positive estimates for  reciprocity, repetition and transitive closure, while a negative one for cyclic closure. The global test \eqref{eq:omnibus_test} is rejected, which leads us to evaluate whether each of the four covariates have been properly modeled, relying on the computation of the one-dimensional statistics in \eqref{eq:KS}. For each of the effects the p-values are all smaller than $\alpha=0.05$, suggesting that all elements in the model are incorrectly included. 
	
	Within this context, we have however access to the timestamps of the events. Consequently, we may choose to evaluate more sophisticated versions of the covariates, specifically those integrating temporal information. Potential model formulations become more diverse, even under the assumption that each relational dynamic is exclusively modelled either as an identity or as a linear function of time. Model selection can be conducted by assessing the AIC, as outlined by \cite{wit2012}. The best model in terms of AIC is given by $\lambda_{sr}(t) = Y_{sr}(t) \times \lambda_0(t) \times \exp{\left[ f_{sr}(t) \right]}$, where
		\begin{equation}\label{eq:mod-appl-time}
			f_{sr}(t) = \beta_{\textmd{rec}} \textmd{rec}^{\textmd{time}}_{sr}(t) + \beta_{\textmd{rep}}  \textmd{rep}^{\textmd{id}}_{sr}(t) + \beta_{\textmd{trs}}  \textmd{trs}^{\textmd{time}}_{sr}(t) + \beta_{\textmd{cyc}}  \textmd{cyc}^{\textmd{id}}_{sr}(t) 
		\end{equation}
	However, the global and individual GOF tests still reject the overall model, as well as each term individually at the $\alpha=0.05$ level. We thus need to look for an alternative and richer model formulations with the information at our disposal.
		
\begin{figure}	
		\begin{tabular}{cc}
			\includegraphics[width=0.50\linewidth]{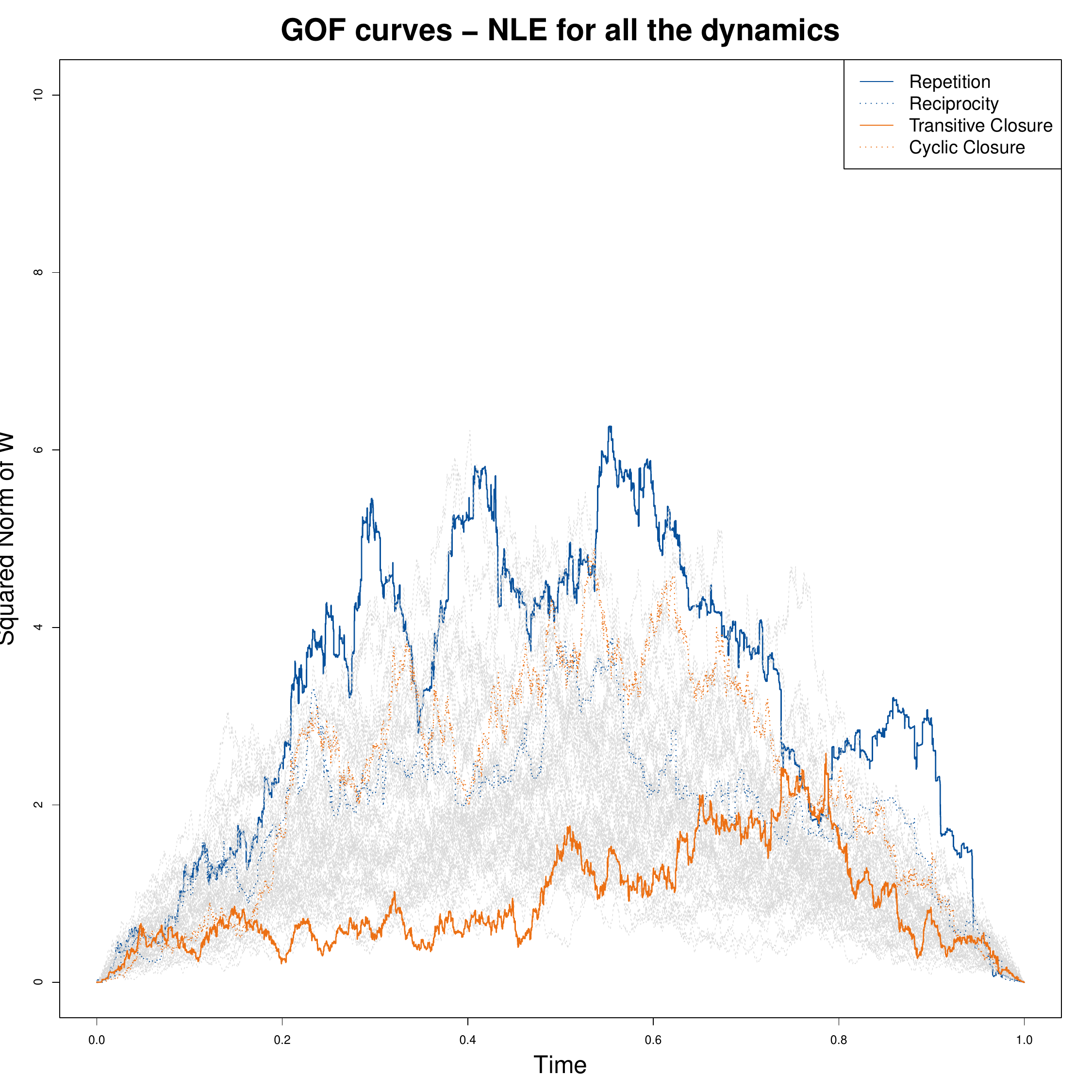} & 
			\includegraphics[width=0.48\linewidth]{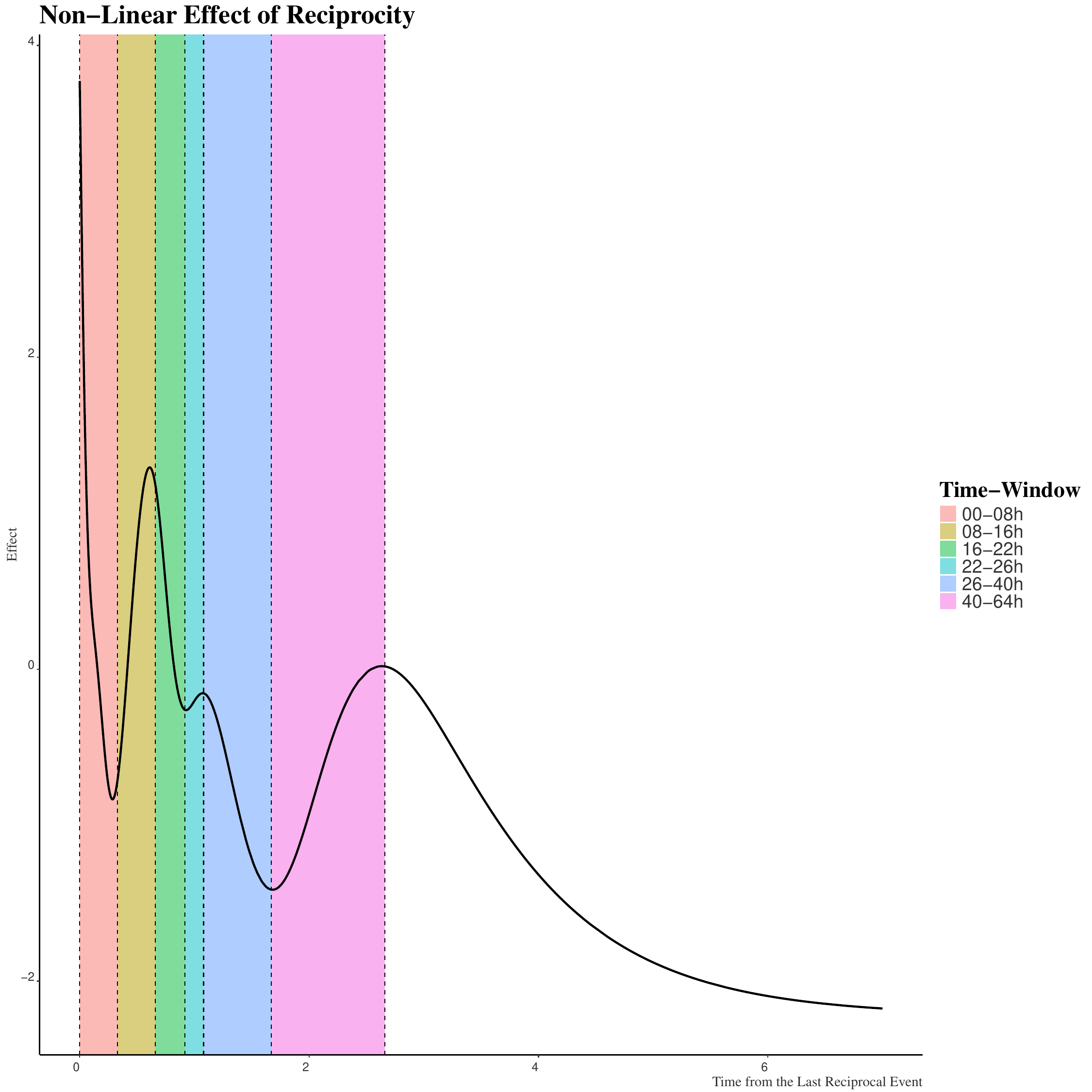}\\
			(a) & (b)
		\end{tabular}	
		\caption{\emph{Left}. GOF process $\hat{\bm{W}}[\hat{\bm{\gamma}}, \cdot]$ evaluated for each of the four endogenous covariates included in model  \eqref{eq:mod-appl-time-re-all-time} with non-linear effects. Although the global test does not find any major evidence against the adequacy of the model, it might be that repetition is still not fully accurately incorporated in the model. It might be possible by changing the time-scale parameter $b$ in \eqref{eq:timescale}. \emph{Right}. Non-linear impact of reciprocity on the intensity of email occurrence. Reciprocity is represented here by the corresponding inter-arrival times of emails. Emails have a tendency to be answered quickly, but this is followed by a steep decline. There is another peak after 16 hours, probably corresponding to the morning after receiving the email the day before. To enhance readability, the background is shaded according to working hour blocks, aligning with typical working and resting hours.}
		\label{fig:application-results}  
	\end{figure}

	Following \cite{juozaitiene2024nodal}, it might be relevant to include in the model formulation random effects accounting for the sender activity and the receiver popularity. In this way, we can effectively model the individual heterogeneity to send (or receive) emails. Furthermore, we consider a non-linear effect for all four temporal versions of the endogenous covariates, i.e.,
	\begin{equation}\label{eq:mod-appl-time-re-all-time}
	 f_{sr}(t) =  f_{1}(\textmd{rec}^{\textmd{time}}_{sr}(t)) +  f_{2}(\textmd{rep}^{\textmd{time}}_{sr}(t)) + 
	 f_{3}(\textmd{trs}^{\textmd{time}}_{sr}(t)) + 
	 f_{4}(\textmd{cyc}^{\textmd{time}}_{sr}(t)) + 
	b^{\textmd{act}}_s + b^{\textmd{pop}}_r 
	\end{equation}
	Each of the endogenous variables are fitted as thin plate regression splines. The random effects are tested as non-properly modeled, but their inclusion leads to an improvement of the four covariates that are globally tested as adequate   (p-value = $0.07$). This is consistent with the discussion in \cite{juozaitiene2024nodal}. Figure~\ref{fig:application-results}a reports the GOF curves $\hat{\bm{W}}[\hat{\bm{\gamma}}, u]$ for the individual model components over the normalized time $u \in [0,1]$. The individual test for repetition has a rather low p-value ($0.019$), suggesting that the fit is not optimal. We have not optimized the temporal scale parameter in \eqref{eq:timescale}, and simply selected $b=1$. By adapting this the model could possibly be further improved.
	
	\subsection{Final model interpretation}
	
	Model interpretation is not the core focus of this paper. Nevertheless, the leading idea of this work is that one should exclusively interpret those model formulations that are assessed as having an adequate model fit. In this case, we focus on the interpretation of the model in \eqref{eq:mod-appl-time-re-all-time}, even if an improvement is still required in terms of the modeling of random effects. Specifically, we focus here on reciprocity. The other non-linear endogenous effects can be found in the Supplementary Materials.
		
	Figure~\ref{fig:application-results}b depicts the estimated function for reciprocity plotted against the time from the last observed reciprocal event. We highlight time from last reciprocal event in a number of intervals. It can be seen that once an email is sent, the likelihood of receiving an answer decreases quickly as time passes during the remainder of the same workday (around $8$ hours). After this minimum, the likelihood of receiving an answer has another peak at $16$ hours after the email, likely because people postponed the response to the following day. A similar pattern can be observed after two workdays following the email, although the increase after the minimum is smaller in absolute value. Around $64$ hours after the email, the chance of receiving an answer continues to decrease without any further increase.
		
	The model includes random sender and receiver effects. The estimated standard deviations are $\hat{\sigma}_{\textmd{act}} = 0.97$ and $\hat{\sigma}_{\textmd{pop}} = 1.54$, suggesting that particularly the receiver has a strong influence on whether a particular event happens. It would be interesting to cross-list this sender and receiver heterogeneity with some kind of exogenous information concerning the actors involved. Unfortunately, information reported in \cite{michalski2014} does not provide obvious clues about the interpretation of the random effects. Furthermore, the rejection of the null of GOF tests for these effects does not allow us to interpret them.
		
	While acknowledging that our final model is not exhaustive, we have presented a computationally efficient method for assessing the adequacy of several potential relational event models that may be fitted to these relational data. There remains the possibility of achieving a more adequate model formulation by appropriately incorporating additional causal determinants.
	
	\section{Discussion}
	
	As has been clear from the empirical analysis there is an intricate relationship between \textit{model fit} and \textit{model adequacy}. In the context of assessing the goodness of fit for relational event models, some literature still relies on \textit{likelihood-based information criteria}, such as AIC or BIC. These methods, while useful for selecting the most predictive or best fitting model from a set of potential candidates, primarily focus on explained variation rather than the adequacy of fit \citep{schecter2018step, juozaitiene2024nodal}.
		
	\textit{Model selection} is a crucial step in relational event modeling, where the potential dynamics influencing event occurrences can be numerous and complex. For instance, \cite{bianchi2023ties} highlights a significant number of statistical measures that can be incorporated into a REM. Likelihood-based methods can be employed to compare these various options effectively. Additionally, model selection is essential for evaluating the importance of including random effects, which aim to explain the intrinsic characteristics of the actors or event types. This aspect is particularly emphasized in \cite{juozaitiene2024nodal}. Nevertheless, selecting the best model based on some information criteria or feature selection method does not guarantee that the model is adequate. We provided an example of this phenomenon in our application section. Model adequacy is crucial because a model that is not adequate cannot be interpreted. 
		
	In practice, any goodness of fit procedure should commence from the global omnibus test. If this test is rejected, the next step is to inspect the individual model components. This inspection can identify where the problem in the model formulation occurred. The temporal structure of the martingale residual process gives the analyst a lead to identify where the problem occurs. In this case, they can examine the specific time frame where the problem arises and speculate on potential causes and improvements. 

	\section{Conclusions}	

	This paper has introduced a comprehensive method for assessing the goodness of fit of relational event models, including their individual components. We propose a procedure that can be easily applied after fitting the relational event model, without requiring additional assumptions about what needs to be tested. This approach focuses on evaluating the actual components of the model formulation. Furthermore, the method can be extended to test for the GOF of arbitrary auxiliary statistics. Therefore, we believe our method can be directly compared to that described by \cite{amati2024goodness}. Our approach, however, offers the possibility of not needing to simulate the entire DGP according to the fitted relational event model, as it is based on computationally more efficient approaches.
		
	Indeed, the advantage of our method, compared to existing GOF techniques, lies in its low computational overhead, particularly when testing a component included in the model formulation. Generally speaking, the primary computational cost involves simulating the theoretical asymptotic behaviour of the process. This is significantly less demanding than the computational cost required to simulate events such as in \cite{brandenberger2019} or \cite{amati2024goodness}, where the simulation of events requires the computation and the updating of endogenous covariates for all potential events (proportional to the square of the number of actors) at each time point of interest.
		
	When one does not reject the GOF of a component of a model, we will usually pretend that the corresponding element is properly specified. However, there might still be misspecifications. It could be that the misspecification is smaller than can be detected by the available data. Or, it could be that omitted confounders might affect the structure and adequacy of the included model component. This is an important consideration for both testing and interpreting the model. For example, in our application study, we interpret the reciprocity effect by referring to the impact of the work-time may have on event patterns. We did not explicitly include the time-of-day in the model formulation (as it is part of the baseline hazard), resulting in the fact that it now "appears'' in the other effects that are included in the model. 

	\section{List of Abbreviations}
	\begin{spacing}{1.5}
		\begin{enumerate}
			\item REM: relational event model;
			\item GOF: goodness of fit;
			\item FLE: fixed linear effect;
			\item NLE: non-linear effect;
			\item TVE: time-varying effect;
			\item RE: random effect;
			\item PL: partial likelihood;
			\item NCC: nested case control;
			\item PMLE: penalized maximum likelihood estimator;
			\item DGP: data generation process;
			\item CS: correctly specified;
			\item MS: misspecified;
		\end{enumerate}
	\end{spacing}
	
	\section{Funding}
	This work was supported by funding from the Swiss National Science Foundation (grant 192549).
	
	\section{Authors' contributions}
	MB and EW designed the study and developed the methodology. MB performed the computational analyses in the manuscript. MB wrote the manuscript with contributions from EW. Both authors read and approved the manuscript. 
	
	\section{Competing Interests}
	The authors declare that they have no competing interests.
	
	\section{Acknowledgments}
	Not applicable. 

	\bibliographystyle{acm}
	\bibliography{gof-bibliography}
	
\end{document}